\documentclass[aps,amsmath,amssymb,floatfix,showpacs,letterpaper,nofootinbib]{revtex4}
\usepackage{graphicx}
\usepackage{enumerate}

\usepackage[figuresleft]{rotating}
\usepackage{color}

\def\bwt{\begin{widetext}}
\def\ewt{\end{widetext}}
\def\be{\begin{equation}}
\def\ee{\end{equation}}
\def\bea{\begin{eqnarray}}
\def\eea{\end{eqnarray}}
\def\bean{\begin{eqnarray*}}
\def\eean{\end{eqnarray*}}
\def\bary{\begin{array}}
\def\eary{\end{array}}
\def\bit{\begin{itemize}}
\def\eit{\end{itemize}}

\def\su5u1{SU(5) \times U(1)}
\def\fsu5u1{SU(5) \times U(1)'}
\def\so10{SO(10)}
\def\sq20{SO(10) \times SO(10)}

\begin{document}

%\iffalse
%\title{Distinguishing the Right-Handed Up/Charm Quarks
%from Top Quark via Discrete Symmetries in the Standard Model Extensions }

\title{$E_6$ GUT through Effects of Dimension-5 Operators }

\author{Chao-Shang Huang$^a$, Wen-Jun Li$^b$, and Xiao-Hong Wu$^c$}

\affiliation{
$^a$ Key Laboratory of Theoretical Physics,
 Institute of Theoretical Physics, Chinese Academy of Sciences,
Beijing 100190, P. R. China \\
$^b$ Institute of Theoretical Physics,
College of Physics and Material Science,
Henan Normal University, Xinxiang 453007, China \\
$^c$ Institute of Modern Physics, School of Science,
East China University of Science and Technology, Meilong Road 130,
Shanghai 200237, China
}

\date{\today}

\begin{abstract}
%\abstract{
In the effective field theory framework, quantum gravity can induce
effective dimension-5 operators,
which have important impacts on grand unified theories.
Interestingly, one of main effects
is
the modification of the usual gauge coupling unification condition.
We investigate the gauge coupling unification in $E_{6}$
under modified gauge coupling unification condition
at scales $M_X$ in the presence of one or more dimension-5 operators.
It is shown that nonsupersymmetric models of $E_6$ unification
can be obtained and can easily satisfy the constraints
from the proton lifetime.
For constructing these models,
we consider several maximal subgroups
$H=SO(10)\times U(1), H=SU(3)\times SU(3)\times SU(3)$,
and $H=SU(2)\times SU(6)$ of $E_{6}$ and
the usual breaking chains for a specific maximal subgroup,
and derive all of the Clebsch-Gordan coefficients $\Phi^{(r)}_{s,z}$
associated with $E_6$ breaking to the Standard Model,
which are given in Appendix A.

%}
\end{abstract}

%\pacs{11.25.Mj, 12.10.Kt, 12.10.-g}

\maketitle

\section{Introduction}
It is well-known that the problem of quantum gravity
has not been solved yet,
although superstring theory presents a beautiful promise.
Can we examine the effects of quantum gravity nowadays?
The answer is positive. The reason is that the scale of unification
$M_{G}\sim 2\times 10^{16}$ GeV
in the minimal supersymmetric Standard Model,
as implicated by the experiment
measurements~\cite{Langacker:1991an,Giunti:1991ta,Amaldi:1991cn,Ellis:1990wk},
is smaller than the Planck scale $M_{Pl}$ ($ M_{Pl}=(8\pi G_{N})^{-1/2}\sim 2.4\times 10^{18}$ GeV),
where quantum gravity should come in,
by about two orders of magnitudes so that
one can build a field theoretical description
of the unification of particle interactions
without a full solution to the problem of quantum gravity.
%That is,
To describe the effects of quantum gravity,
we could use the effective field theory approach
in which non-renormalizable higher dimension operators are introduced.
The $d\geq 5$ operators induced by gravity should enter the Lagrangian,
which are suppressed by factors of $(M_{Pl})^{-(d-4)}$
with coefficients at the order of $\sim {\cal O}(1)$ .
They are only subject to the constraints of the symmetries
(gauge invariance, supersymmetry in supersymmetric models, etc.)
of the low energy theory.

As it has been shown in Refs.~\cite{Hill:1983xh,Shafi:1983gz,
Calmet:2008df,Hall:1992kq,Datta:1995as,Huitu:1999eh,
Dasgupta:1995js,Tobe:2003yj,Nath:2006ut,Rizzo:1984mk,
Marciano:1981un,Anderson:1996bg,Amundson:1996nw,Martin:2009ad,
Balazs:2010ha,Balazs:2011ie,Wang:2011eh,Chakrabortty:2008zk,
Lykken:1993br,Howl:2007hq,Ellis:1985jn,Drees:1985bx,Calmet:2009hp}
that the presence of higher dimension operators may have substantial impacts
on grand unified theory (GUT)
and its phenomenology for gauge groups $SU(5)$ and $SO(10)$.
These operators modify the usual gauge coupling unification
condition~\cite{Hill:1983xh,Shafi:1983gz,Calmet:2008df,Hall:1992kq,
Datta:1995as,Huitu:1999eh,Chakrabortty:2008zk,Calmet:2009hp}.
It is estimated that the effects of dimension-5 operators
can be more important
than two-loop corrections in the renormalization group analysis
of gauge coupling unification~\cite{Calmet:2008df}.
With one or more dimension-5 operators,
it is possible to achieve unification at scales $M_X$
much different than usually expected~\cite{Calmet:2009hp}.
Higher dimension operators can lead to an acceptable value of
$\sin^2\theta_w$~\cite{Hill:1983xh,Rizzo:1984mk,Marciano:1981un,
Chakrabortty:2008zk},
affect supersymmetric (SUSY) particle spectrum in SUSY GUT
and supergravity~\cite{Anderson:1996bg,Amundson:1996nw,Martin:2009ad,
Balazs:2010ha,Balazs:2011ie,Wang:2011eh,Chakrabortty:2008zk,
Lykken:1993br,Howl:2007hq,Ellis:1985jn,Drees:1985bx}
and the analysis of proton decay~\cite{Tobe:2003yj,Nath:2006ut}.
Therefore, we should include them in the researches of GUT and SUSY GUT.
In particular, in some cases, such studies are dramatically important.
For example,
the gauge coupling unification in the minimal SU(5) without supersymmetry can not be realized
and the minimal SUSY SU(5) has been already excluded
by the limit (larger than $10^{34}y$) of the proton lifetime
from Super-Kamiokande~\cite{TheSuper-Kamiokande:2017tit}\footnote{SU(5) GUT
can be obtained by adding more Higgs representations
or discrete symmetries~\cite{Georgi:1979df,FileviezPerez:2007nh,
Arbelaez:2015toa}.}, but they can be realized
if effects of $d\geq 5$ operators are
included (see, for example,~\cite{Chakrabortty:2008zk,Calmet:2009hp}).

The exceptional group $E_{6}$ is
an attractive unification group among well-known unification groups.
The main reasons are as follows.
Firstly, from the viewpoint of superstring theory,
the gauge and gravitational anomaly cancellation
occurs only for the gauge groups $SO(32)$
or $E_{8} \times E_{8}$~\cite{Green:1984sg,Green:1987sp,Green:1987mn}
and compactification on a Calabi-Yau manifold
with an SU(3) holonomy results in the breaking
$E_{8} \rightarrow SU(3) \times E_{6}$~\cite{Candelas:1985en}.
Secondly, in terms of the phenomenology of low energy effective theories
originated from $E_{6}$ GUT, there are several attractive features~\cite{King:2005jy,Kawase:2010na,Athron:2013ipa,Rosner:2014cha,Nevzorov:2013tta,
Nevzorov:2015sha,Rojas:2015tqa,Nevzorov:2015iya}.
Moreover, if we assume the dynamical symmetry breaking scenario,
we would have several constraints on the possible GUT models.
It has been pointed out~\cite{Barbieri:1980tz,Barbieri:1981yy,Ramond:1979py}
that $E_{6}$ is uniquely selected among many GUT groups,
if one demands that 1) the theory is automatically anomaly free,
2) every generation of quark/lepton fields belongs to
a single irreducible representation (irrep) of the GUT group,
and 3) the Higgs fields,
which are necessary for inducing
the symmetry breaking down to $SU_{C} (3)\times U_{\rm em}(1)$,
fall in the representations that can be provided by the fermion bilinears.
Recently some models and their low energy phenomenology
originated from $E_6$ GUT generated further interests~\cite{Athron:2016gor,
Joglekar:2016yap,Nevzorov:2014sha,Dhuria:2015swa,Athron:2015vxg,
Harada:2016fpj,Athron:2016qqb,Belanger:2017vpq,Nevzorov:2017rtf}.

Effects of higher dimension operators to the unification of
gauge couplings have also been investigated in
the grand unified gauge group $E_6$~\cite{Chakrabortty:2008zk,Huang:2014zba}.
In the Ref.~\cite{Chakrabortty:2008zk},
$E6\rightarrow SU(3)\times SU(3)\times SU(3)$  has been studied
and the corresponding effective contributions
to gauge kinetic terms have been given.
However, their contributions to gauge kinetic terms
in the Standard Model (SM) have been not given,
although it is not difficult to derive them from the results
shown in Table 4 of the paper. Moreover, the numerical analysis on the unification of gauge couplings and corresponding physical discussions
have not been carried out in the paper.
% in the paper. %Ref.~\cite{Chakrabortty:2008zk}.
To construct $E_{6}$ unification models
with effects of dimension-5 operators in Ref.~\cite{Huang:2014zba},
the author considers only %the case of
the maximal subgroup $H=SO(10)\times U(1)$ and
gives the Clebsch-Gordan coefficients $\Phi^{(r)}_{s,z}$
associated with $E_6$ breaking to the SM for that case.
Nevertheless, same as the Ref.~\cite{Chakrabortty:2008zk}, the numerical analysis on the unification
of gauge couplings and corresponding physical discussions
have not been carried out.

In this paper, we investigate the unification of gauge couplings
for the grand unified gauge group $E_6$
through effects of dimension-5 operators.
Surveying the branching rules for the GUT group
$E_{6}$ \cite{Slansky:1981yr},
we see that there are several maximal subgroups
containing $G_{321}$ (e.g., $H=SO(10) \times U(1)$,
$H=SU(3)\times SU(3)\times SU(3)$, $H=SU(2)\times SU(6)$, and $H=F_{4}$)
and for a specific maximal subgroup
there are usually several breaking chains.
We consider all common maximal classical subgroups
%$H=SO(10)\times U(1),SU(3)\times
%SU(3)\times SU(3)$ and $H=SU(2)\times SU(6)$,
with $SU_C(3) \times U(1)_{\rm em}$
and the usual breaking chains for a specific maximal subgroup. % for specific one. %In the paper we
%investigate $E_6$ GUT with effects of dimension-5 operators for all maximal %classical subgroups with $U(1)^{em}\times SU_c(3)$.
We show that the gauge coupling unification condition is modified
due to the effects of dimension-5 operators,
which are the lowest higher dimension operators in $E_6$ GUT,
and the gauge coupling unification at a higher scale
can be realized without SUSY.
% for modified gauge coupling unification condition.
Furthermore, we find that the constraints
from the proton lifetime can be safely satisfied
even at the higher unification scale $M_{G}$
near the Planck scale $M_{Pl}$.

In section II, we set up the notation to study effects
of dimension-5 operators and point out
how dimension-5 operators modify
the usual gauge coupling unification condition.
Section III is devoted to the numerical analysis of
renormalization group (RG) evolution of the gauge couplings
and the corresponding physical results for these cases.
In section IV, we discuss the constraints from the proton lifetime.
Summary and conclusions are given in section V.
All of the Clebsch-Gordan coefficients $\Phi^{(r)}_{s,z}$
associated with $E_6$ breaking to the SM,
in different bases $\{s,z\}$, up to a uniform normalization constant
for different representations $r$, are derived, and results are
given in Appendix A.
In Appendix B, we present the structure constants of $E_6$ explicitly,
which are mostly used by physicists and in the study of GUT.

 \section{dimension-5 operators and modified
gauge coupling unification condition}

Dimension-5 operators are singlets of the grand unified gauge group G,
and are formed from gauge field strengths $G_{\mu\nu}$
and Higgs multiplets $H_{k}$ of G,
\be
{\cal L} =\frac{c_{k}}{M_{Pl}} G^{a}_{\mu\nu} G^{b \mu\nu} H_{k}^{ab},
\label{dim5}
\ee
where a,b are group indices and k labels different multiplets.
Therefore, % the kinetic terms of gauge bosons
the terms relevant to our discussions of gauge coupling unification
in the Lagrangion at the unification scale are,
\be
{\cal L} =-\frac{1}{4} G^a_{\mu\nu}G^{a \mu\nu} + \frac{c_{k}}{4 M_{Pl}} G^{a}_{\mu\nu} G^{b \mu\nu} H_{k}^{ab}. \label{Kine}
\ee
Now G=$E_6$, it is evident from Eq. (\ref{dim5})
that the representations, to which Higgs fields $H_{k}$ belong,
can only be contained in the symmetric product of
two adjoints,~\footnote{For simplicity, we use "$\times$" and "+"
to denote the direct product and the direct sum respectively,
whenever there is no confusion.}
\begin{eqnarray}
({\bf 78}\times{\bf 78})_{symmetric}&=&{\bf 1}+{\bf 650}+{\bf 2430}. \label{reps}
\end{eqnarray}
For a specific irrep $r$, $r=1,650,2430$, in Eq. (\ref{reps}),
we denote the Higgs multiplet by a d-dimensional symmetric matrix
$\Phi^{(r)}$, with d=d(G),
the dimension of the adjoint representation (rep) G
(we use the same letter G to denote the group
and its adj. rep. for simplicity.),
and $d=78$ for $G=E_6$.
For our purpose, we find that all possible $\Phi^{(r)}$ are invariant under the SM gauge group
$G_{321}\equiv SU(3)\times SU(2)\times U_{Y}(1)$.
That is, each of them is a SM singlet and then
%in this case
the matrix is largely simplified: it contains only a few independent entries.

There are different ways to define the hypercharge $Y$,
which are consistent with the SM
(see, e.g., the Refs~\cite{Harada:2003sb,London:1986dk}).
We consider two cases:
1) $U_{Y}(1)$ is a subgroup of H;
2) $U_{Y}(1)$ is a subgroup of $E_6$.
Hereafter, we shall call the case 1) as "normal embedding",
and the case 2) as "flipped embedding".
For example, for $H=SU(3)\times SU(3)\times SU(3)$,

\begin{enumerate}%[(i)]
\item normal embedding, $G_{321}\subset SU(3)_C \times SU(2)_L\times SU(2)_R \times U(1)_L \times U(1)_R \subset SU(3)_C\times SU(3)_L\times SU(3)_R \subset E_6$
\begin{eqnarray}\label{e6nomal}
{\bf 78} &\stackrel{SU(3)_C\times SU(3)_L\times SU(3)_R }{\longrightarrow}& \left({\bf 8,1,1}\right) \oplus~~\left({\bf 1,8,1}\right) ~~\oplus~~\left({\bf 1,1,8}\right) \oplus~~\left({\bf 3,\overline 3,\overline{\bf 3}}\right)  \oplus~~\left({\bf \overline 3,3,3 }\right)\nonumber\\
&\stackrel{G_{321}}{\longrightarrow}& \underbrace{({\bf 8},{\bf 1})_0}_{I} \oplus \underbrace{({\bf 1},{\bf 3})_0}_{II} \oplus \underbrace{({\bf 1},{\bf 1})_0}_{III} \oplus \underbrace{\left(({\bf 3},{\bf 2})_{-\frac{5}{6}} \oplus {\rm h.c.}\right)}_{IV}\oplus\underbrace{({\bf 1},{\bf 1})_0}_{V}\nonumber\\
&&~~\oplus~~\underbrace{\left(({\bf 3},{\bf 2})_{\frac{1}{6}} \oplus {\rm h.c.}\right)}_{VI} \oplus \underbrace{\left(({\bf 3},{\bf 1})_{\frac{2}{3}} \oplus {\rm h.c.}\right)}_{VII} \oplus \underbrace{\left(({\bf 1},{\bf 1})_{-1} \oplus {\rm h.c.}\right)}_{VIII} \oplus~~\underbrace{({\bf 1},{\bf 1})_0}_{IX} \nonumber\\
%\newpage
&&~~\oplus \underbrace{\left(({\bf 3},{\bf 2})_{\frac{1}{6}}\oplus {\rm h.c.}\right)}_{X}\oplus \underbrace{\left(({\bf 3},{\bf 1})_{\frac{2}{3}} \oplus {\rm h.c.}\right)}_{XI}
 \oplus \underbrace{\left(({\bf 1},{\bf 1})_{1} \oplus {\rm h.c.}\right)}_{XII} \nonumber\\
 &&~~\oplus  \underbrace{\left(({\bf 3},{\bf 1})_{-\frac{1}{3}} \oplus {\rm h.c.}\right)}_{XIII}\oplus \underbrace{\left(({\bf 1},{\bf 2})_{\frac{1}{2}} \oplus {\rm h.c.}\right)}_{XIV}\oplus \underbrace{\left(({\bf 1},{\bf 1})_{0} \oplus {\rm h.c.}\right)}_{XV}.
\end{eqnarray}
\item flipped embedding, $G_{321}\subset SU(3)_C \times SU(2)_L\times SU(2)_R \times U(1)_L \times U(1)_R \subset SU(3)_C\times SU(3)_L\times SU(3)_R \subset E_6$
\begin{eqnarray}\label{3nomal}
{\bf 78}
&\stackrel{G_{321}}{\longrightarrow}& \underbrace{({\bf 8},{\bf 1})_0}_{I} \oplus \underbrace{({\bf 1},{\bf 3})_0}_{II} \oplus \underbrace{({\bf 1},{\bf 1})_0}_{III} \oplus \underbrace{\left(({\bf 3},{\bf 2})_{\frac{1}{6}} \oplus {\rm h.c.}\right)}_{IV}\oplus\underbrace{({\bf 1},{\bf 1})_0}_{V}\nonumber\\
&&~~\oplus~~\underbrace{\left(({\bf 3},{\bf 2})_{\frac{1}{6}} \oplus {\rm h.c.}\right)}_{VI} \oplus \underbrace{\left(({\bf 3},{\bf 1})_{-\frac{1}{3}} \oplus {\rm h.c.}\right)}_{VII} \oplus \underbrace{\left(({\bf 1},{\bf 1})_{0} \oplus {\rm h.c.}\right)}_{VIII} \oplus~~\underbrace{({\bf 1},{\bf 1})_0}_{IX} \nonumber\\
%\newpage
&&~~\oplus \underbrace{\left(({\bf 3},{\bf 2})_{-\frac{5}{6}}\oplus {\rm h.c.}\right)}_{X}\oplus \underbrace{\left(({\bf 3},{\bf 1})_{\frac{2}{3}} \oplus {\rm h.c.}\right)}_{XI}
 \oplus \underbrace{\left(({\bf 1},{\bf 1})_{-1} \oplus {\rm h.c.}\right)}_{XII} \nonumber\\
 &&~~\oplus  \underbrace{\left(({\bf 3},{\bf 1})_{\frac{2}{3}} \oplus {\rm h.c.}\right)}_{XIII}\oplus \underbrace{\left(({\bf 1},{\bf 2})_{\frac{1}{2}} \oplus {\rm h.c.}\right)}_{XIV}\oplus \underbrace{\left(({\bf 1},{\bf 1})_{-1} \oplus {\rm h.c.}\right)}_{XV}.
\end{eqnarray}
\end{enumerate}

Therefore, $\Phi^{(r)}$, which are invariant under $G_{321}$,
can be written as,
\bea  \label{phir}
(\Phi^{(r)})^{ab}&=& \delta^{ab}~diag[h1~I_8,h2~I_3,h3,h4~I_{12},h5,h6~I_{12},h7~I_6,h8~I_2,h9,h10~I_{12},h11~I_6,h12~I_2,\nonumber\\
&&~~h13~I_6,h14~I_4,h15~I_2]+\delta^{a12}\delta^{b25} h35+\delta^{a~25}\delta^{b~12} h53+\delta^{a~12}\delta^{b~46}h39\nonumber\\
&&~~+\delta^{a~25}\delta^{b~46} h59+\delta^{a~46}\delta^{b~12} h93+\delta^{a~46}\delta^{b~25} h95,
\eea
where $I_n$ is n-dimensional unit matrix,
and $hij=hji$, since $\Phi^{(r)}$ is the symmetric matrix.
We have ordered indices $a, b$ according to the order of terms
in Eq. (\ref{e6nomal}), i.e., $a=1,...,8$ for $h1$,
$a=9,10,11$ for $h2$, $a=12$ for $h3$, ..., et al.
Because $hj$ ($j=1,2,...,15$)
%(for the meaning of $hj$, see next %section)
are the same for both the normal and flipped embedding,
except $h3, h5, h9$, we denote the different entries
in the case of flipped embedding by $h3', h5', h9'$ and $hij'$.

In order to find $\Phi^{(r)}$,
we use the second order Casimir operator,
\be \label{casimir}
C_R=-\sum_{i=1}^{d(G)} X_i^2,
\ee
where R is a irrep of G, and $X_i$'s are the generators of G,
which satisfy
\be \label{afsc}
[X_i,X_j]= i~f_{ijk} X_k
\ee
with $f_{ijk}$ as the totally antisymmetric structure constants.
The operator acts on the tensor product $R\times R$ so that
\be
C_{R\times R}= C_R\times {\bf {\it 1}} + {\bf {\it 1}} \times C_R + 2 F, \ee where
\be \label{ff}
F= -\sum_{i=1}^{d(G)} X_i \times X_i .
\ee
It is easy to derive $F \Phi^{(r)}= F_{G}(r) \Phi^{(r)}$,
where $F_{G}(r)=C(r)/2-C(G)$ is the eigenvalue of $F$
in irrep r with $C(r)$ and $C(G)$ being the eigenvalues of the Casimir operator in irrep r and G respectively.
$C(r)$ depends on the normalization of the Casimir operator
and consequently the choice of structure constants~\footnote{The choice
in this paper leads to that for the subgroup SO(10) of E6,
$c(r)'$s are the same as those in refs.~\cite{Chamoun:2001in,
Chamoun:2009nd,Calmet:2009hp,Girardi:1980um}.}.
The structure constants of $E_6$ have been given in a Chevalley base~\cite{sc}.
For convenience of the study in GUT, they are transformed into the usual form and listed in Appendix B.
%and choose them such that the part %corresponding to $SO(10)$ is the same as %that in the Ref.~\cite{chlw,chr,gss}.

The eigenvalues $C(r)$ for several irreps r in $E_6$,
as well as $SO(10)$ and $SU(6)$ are listed in Table I.

\begin{table}[!h]
\begin{center}
\begin{tabular}{ccccccccc}%{l||c|c|c|c|c|c}
\hline
$E_6$& r &~ 27 &~ 78 &~ 650 &~ 2430 \\
\hline
 & $ C(r)$ &~ 3 &~ 12 &~ 18 &~ 26   \\
\hline
\hline
$SO(10)$ & r & 10 & 45 & 54 & 210 & 770    \\
\hline
 & $ C(r)$ & 1 & 8 & 10 & 12 & 18    \\
\hline
\hline
$SU(6)$ & r &~ 6 &~ 35 &~ 189 &~ 405     \\
\hline
 & $ C(r)$ &~ 1/2 &~ 6 &~ 10 &~ 14    \\
\hline
\end{tabular}
\end{center}
\caption{Quadratic Casimir invariants $C(r)$ for some irreps
$r$ of $E_6$,  $SO(10)$ and $SU(6)$,
in the conventions explained in the section II of the text.
$C({\bf 1})=0$ for the singlet ${\bf 1}$ of any group.\label{C(r)table}}
\end{table}

In the case of $H=SU(6)\times SU(2)$,
depending on different embeddings of the SM into H,
i.e., different specific assignment of the chiral fields
to representations of subgroup H,
there are three cases.
In the first case, denoted as H=$SU(6)\times SU(2)_X$,
the SM is totally in $SU(6)$ and no relation
with the factor group $SU(2)$ of H.
In the second case, denoted as H=$SU(6)\times SU(2)_L$,
$SU(2)_L$ is the factor group $SU(2)$ of H.
And in the third case, denoted as H=$SU(6)\times SU(2)_R$,
$SU(2)_R$ is the factor group $SU(2)$ of H.
For all the cases, % we considered,
a direct but tedious calculation gives all of the Clebsch-Gordan
coefficients, $\Phi^{(r)}_{s,z}$, associated with $E_6$ breaking
to the SM, which are listed in Appendix A.

By classifying the SM singlets $\Phi_{s,z}^{(r)}$ into irreps
%according to their transformation properties
under different breaking chains as in Tables in Appendix A,
we do not mean to imply a grand unified symmetry breaking scenario,
where $E_6$ is broken to the SM necessarily
via one or some intermediate gauge groups,
though such a classification is well suited for such a scenario.
However, such a classification is suitable to serve
as a parametrization for the SM singlets $ <H_k^{ab}> $,
the nonzero vacuum expectation values of $H_k^{ab}$,
transforming in irrep $r_k$~\footnote{The convention is:
k=1,2,3 correspond to irreps ${\bf 1},{\bf 650},{\bf 2430}$.} of $E_6$,
in terms of the basis $\Phi^{k}_{s,z}\equiv \Phi^{(r_k)}_{s,z}$,
\be \label{vavs}
<H_k^{ab}> \equiv \sum_{s,z} v_{s,z}^{k}~ \Phi^{k~ab}_{s,z}\equiv v^k~\Phi^{k~ ab}, \ee
where $v_{s,z}^{k}$, $v^k$ are real.
Eqs.~(\ref{Kine}), (\ref{vavs}) lead to an alteration of
the gauge coupling unification condition,
\be \label{gaucon}
g^2_1(M_{GUT})(1 + \epsilon_1) = g^2_2(M_{GUT})(1 + \epsilon_2) = g^2_3(M_{GUT})(1 + \epsilon_3)=g^2_G/4\pi, \ee
where
\be \label{epsilond}
\epsilon_i = \sum_k \frac{c_k}{M_{Pl}} \sum_{s,z} v_{s,z}^k \phi_i^{k~sz}, ~~~i = 1, 2, 3, ~~~~ %\nonumber \\
 \phi_1^{k~sz}\equiv -h3^{k}_{s,z},~~\phi_2^{k~sz}\equiv -h2^{k}_{s,z},~~\phi_3^{k~sz}\equiv -h1^{k}_{s,z}, \ee
for the normal embedding,
and $\phi_i^{k~sz}$ are listed in Tables (see Appendix A).
For the case of flipped embedding, $ \phi_1^{k~sz}$
is changed into $\phi_1^{k~sz}\equiv -h3'^{k}_{s,z}$.
$h3'^{k}_{s,z}$ are also listed in Tables in Appendix A.

Eq.~(\ref{gaucon}) is the boundary condition at the scale $M_{GUT}$
of the RG evolution of gauge couplings,
and will be used in the numerical analysis in the next section.

\section{Grand unification in $E_6$}

In this section, we study the unification of gauge couplings
with one, two, or more dimension-5 operators numerically
from different $\bf 650$ and $\bf 2430$ breaking chains.

As to the case of one dimension-5 operator from ${\bf 650}$ or ${\bf 2430}$, because there are several maximal subgroups and
for a specific maximal subgroup there are several breaking chains, the SM singlets $ <H_k^{ab}> $,
the nonzero vacuum expectation values of $H_k^{ab}$,
are not determined for fixed k=2 or 3.
Consequently it is clear from Eq. (\ref{epsilond}) and
Tables 3-10 that the ratio $\epsilon_1 : \epsilon_2 : \epsilon_3$
can not be determined fully.
So one has much freedom to choose ratios among $v_{s,z}^{k}$.
It is a boring and not necessary task to exhaust all possibilities.
Instead, we just take some breaking chains
%and assume some ratios among $v_{s,z}^{k}$
as examples.
Then, the unification scale $M_X$
(set $M_X=M_{GUT}$ for simplicity) and Wilson coefficient $c_k$
are computable by means of the gauge coupling unification
condition Eq.(\ref{gaucon}), when the running gauge couplings
in the SM are given. We limit ourselves to one-loop case for running
in the paper for simplicity and it is straightforward
to generalize to two-loop case.
Moreover, dimension-5 operators also
affect analysis of proton decay~\cite{Tobe:2003yj,Nath:2006ut}.
As analyzed in the Ref.~\cite{Calmet:2009hp} and in the next section,
the absolute value of Wilson coefficient should be less than 10,
i.e., max$|c_k| \le 10$, in order to satisfy the proton decay constraint.
Hereafter, we name the unification with max$|c_k| \le 10$
as the successful unification.

%\clearpage
\begin{table}[ht]
%\begin{center}
\tabcolsep=0.04cm
\footnotesize{
\begin{tabular}{c|c||c|c|c|c|c|c|c|c|c|c|c}
\hline
E$_6$ embedding & breaking chain & $M_X$(GeV) & $v$(GeV) & $c_k$ &
 $1/\alpha_G$ & $\epsilon_1$ & $\epsilon_2$ & $\epsilon_3$ \\
\hline
$SU(5)\subset SO(10)$ & ${\bf 650} \to {\bf 54} \to {\bf 24}$ &
 $4.0\times 10^{13}$ & $9.8\times 10^{13}$ & $1.1\times 10^3$ &
 $40.6$ & $0.023$ & $0.068$ & $-0.045$ \\
%\hline
                      & ${\bf 2430} \to {\bf 210} \to {\bf 75}$ &
 $3.7\times 10^{15}$ & $7.8\times 10^{15}$ & $-6.4$ &
 $43.0$ & $-0.103$ & $0.062$ & $0.021$ \\
\hline
$SU(3)_L \subset SU(6)\times SU(2)_X$ &
 ${\bf 650} \to ({\bf 35, 1}) \to {\bf 1}$ &
 $4.0\times 10^{13}$ & $9.8\times 10^{13}$ & $1.4\times 10^3$ &
 $41.1$ & $0.011$ & $0.056$ & $-0.056$ \\
    & ${\bf 2430} \to ({\bf 405, 1}) \to {\bf 1}$ &
 $6.4\times 10^{16}$ & $1.4\times 10^{17}$ & $0.63$ &
 $48.9$ & $-0.250$ & $-0.038$ & $-0.038$ \\
\hline
$SU(3)_L \subset SU(6)\times SU(2)_L$ &
    ${\bf 2430} \to ({\bf 189, 1}) \to {\bf 1}$ &
 $3.6\times 10^{18}$ & $8.1\times 10^{18}$ & $-0.015$ &
 $49.0$ & $-0.306$ & $0.$ & $0.051$ \\
\hline
\end{tabular}
}
%\end{center}
\caption{\label{single}
The breaking chain, unification scale $M_X$, Higgs VEV $v$, Wilson coefficient $c_k$, gauge coupling $\alpha_G$ at the unification scale,
and three $\epsilon_{1,2,3}$ for unification
with only one dimension-5 operator.
}
\end{table}

\begin{figure}[ht]
\centerline{ \includegraphics[height=8cm,width=7cm]{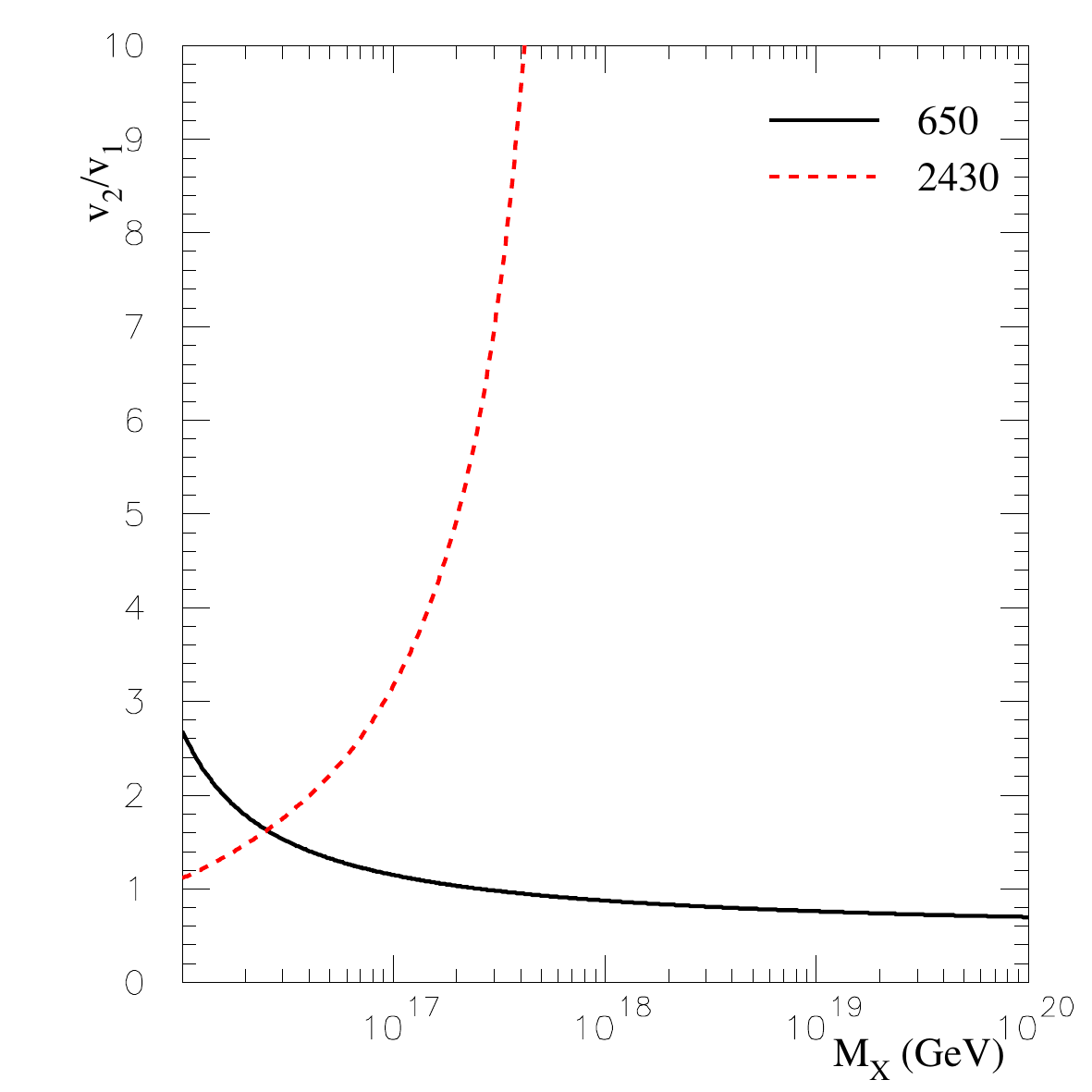} }
\caption{ \label{so10same}
The ratio of VEVs as a function of unification scale $M_X$
according to the transformation properties
under the subgroup embedding $SU(5)\subset SO(10) \subset E_6 $,
for two breaking chains, ${\bf 210} \to {\bf 24}$ and
${\bf 210} \to {\bf 75}$ with ${\bf 650}$ (solid line), and,
two breaking chains, ${\bf 770} \to {\bf 24}$ and
${\bf 770} \to {\bf 200}$ with ${\bf 2430}$ (dashed line).
The ratio $v_2/v_1$ as $v_{\bf 75}/v_{\bf 24}$ for ${\bf 650}$,
and $v_{\bf 200}/v_{\bf 24}$ for ${\bf 2430}$.
}
\end{figure}

\begin{figure}[htb]
\centerline{ \includegraphics[height=8cm,width=7cm]{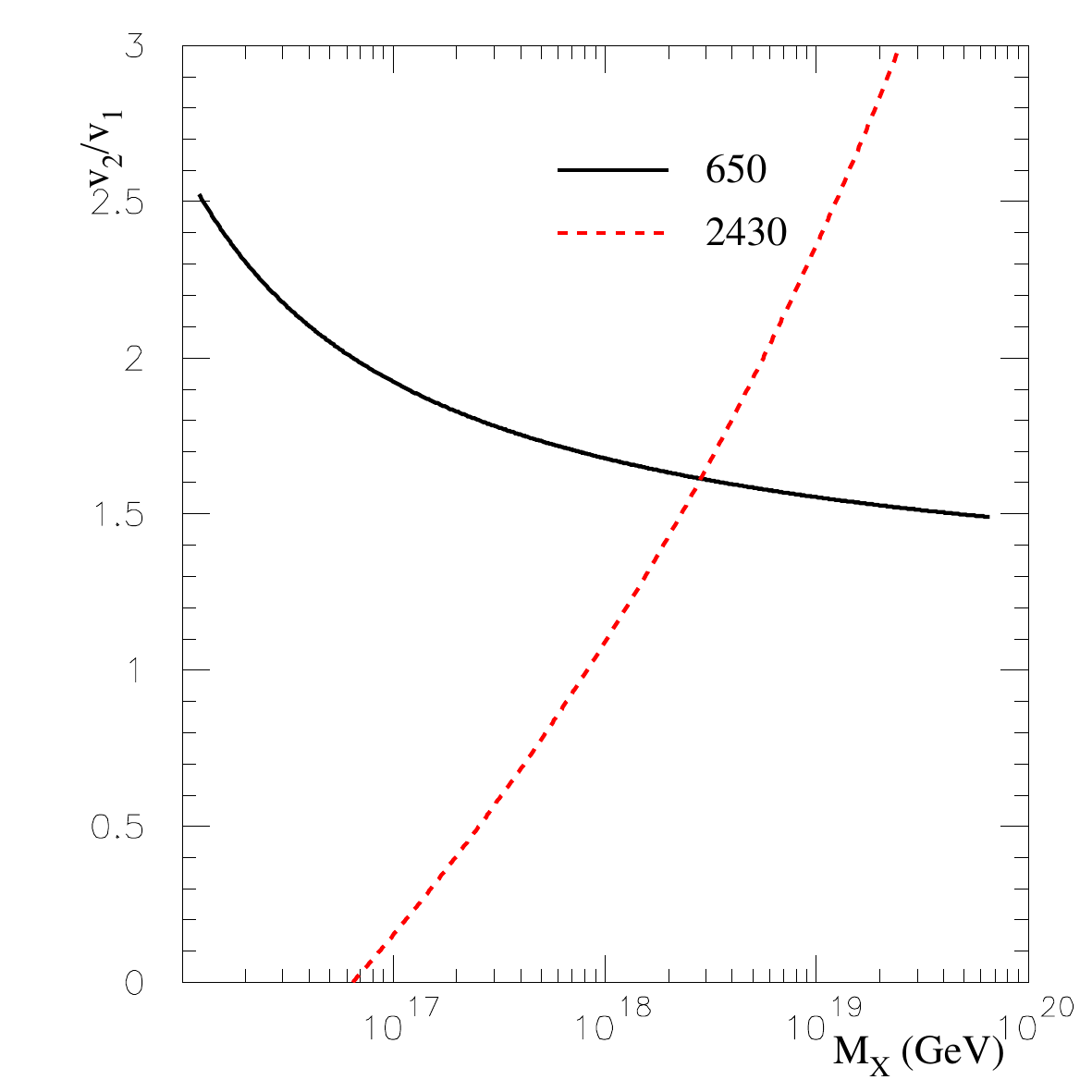} }
\caption{ \label{Xsame}
The ratio of VEVs as a function of unification scale $M_X$
according to the transformation properties
under the subgroup embedding $SU(3)_L \subset SU(6) \times SU(2)_X \subset E_6$,
for two breaking chains, ${\bf (35,1)} \to {\bf 1}$ and
${\bf (189,1)} \to {\bf 8}$ with ${\bf 650}$ (solid line), and,
two breaking chains, ${\bf (405,1)} \to {\bf 1}$ and
${\bf (405,1)} \to {\bf 8}$ with ${\bf 2430}$ (dashed line).
The ratio $v_2/v_1$ as $v_{\bf 8}/v_{\bf 1}$
for both ${\bf 650}$ and ${\bf 2430}$.
}
\end{figure}

In Tab.~\ref{single}, with only one breaking chain, five example results
are given with their respective
unification scale $M_X$, Higgs VEV $v$, Wilson coefficient $c_k$,
gauge coupling $\alpha_G$ at the unification scale,
and three $\epsilon_{1,2,3}$ for the unification.
The needed Wilson coefficients are too large in two cases of
${\bf 650} \to {\bf 54} \to {\bf 24}$ and
${\bf 650} \to ({\bf 35, 1}) \to {\bf 1}$.
While in the other three cases of ${\bf 2430} \to {\bf 210} \to {\bf 75}$,
${\bf 2430} \to ({\bf 405, 1}) \to {\bf 1}$,
and ${\bf 2430} \to ({\bf 189, 1}) \to {\bf 1}$,
we can reach the successful unification.
In particular, for ${\bf 2430} \to ({\bf 189, 1}) \to {\bf 1}$,
$M_X$ is as large as the Planck scale
and $|c_k|$ is as small as $-0.015$.
That means, we may have a perturbative theory
up to the onset of quantum gravity.
With only one dimension-5 operator,
we can also use two or more different breaking chains
to achieve successful gauge coupling unification
with continuously varied unification scale $M_X$,
as given in Fig.~\ref{so10same},~\ref{Xsame} for illustrations.
In Fig.~\ref{so10same}, we choose two breaking chains
${\bf 210} \to {\bf 24}$ and
${\bf 210} \to {\bf 75}$ for ${\bf 650}$,
${\bf 770} \to {\bf 24}$ and
${\bf 770} \to {\bf 200}$ for ${\bf 2430}$, respectively,
under the subgroup embedding $SU(5)\subset SO(10) \subset E_6 $.
We plot the ratios of VEVs, $v_{\bf 75}/v_{\bf 24}$ for ${\bf 650}$,
and $v_{\bf 200}/v_{\bf 24}$ for ${\bf 2430}$
as a function of continuously varied $M_X$ with
successful unification.
%where absolute value of Wilson coefficient less than 10, max$|c_k| \le 10$}.
In Fig.~\ref{Xsame}, the ratios of VEVs,
$v_{\bf 8}/v_{\bf 1}$ for both ${\bf 650}$ and ${\bf 2430}$,
as a function of $M_X$ are plotted,
with ${\bf (35,1)} \to {\bf 1}$ and
${\bf (189,1)} \to {\bf 8}$ for ${\bf 650}$, and,
${\bf (405,1)} \to {\bf 1}$ and
${\bf (405,1)} \to {\bf 8}$ for ${\bf 2430}$,
under the subgroup embedding
$SU(3)_L \subset SU(6) \times SU(2)_X \subset E_6$.
As it is easy to see, from Eqs.~(\ref{gaucon}), (\ref{epsilond})
and Tables in Appendix I and in the Ref~\cite{Huang:2014zba}, that for a given specific breaking chain,
when two of three numbers
$\{-h1^{k}_{s,z}, -h2^{k}_{s,z},-h3^{k}_{s,z}(h3'^{k}_{s,z})\}$ are zero,
i.e., for $(45,0)$, $(210,0)\to {\bf 1}$ in $H=SO(10)\times U(1)$,
$(189,1)$, $(405,1)\to {\bf 8,27}$ in $H=SU(6)\times SU(2)_L$,
$(35,3), (1,5)$ in $H=SU(6)\times SU(2)_R$ and $(1,8), (8,8), (1,27)$
in $H=SU(3)\times SU(3)\times SU(3)$,
one can not get successful unification.

\begin{figure}[ht]
\centerline{ \includegraphics[height=7cm,width=6cm]{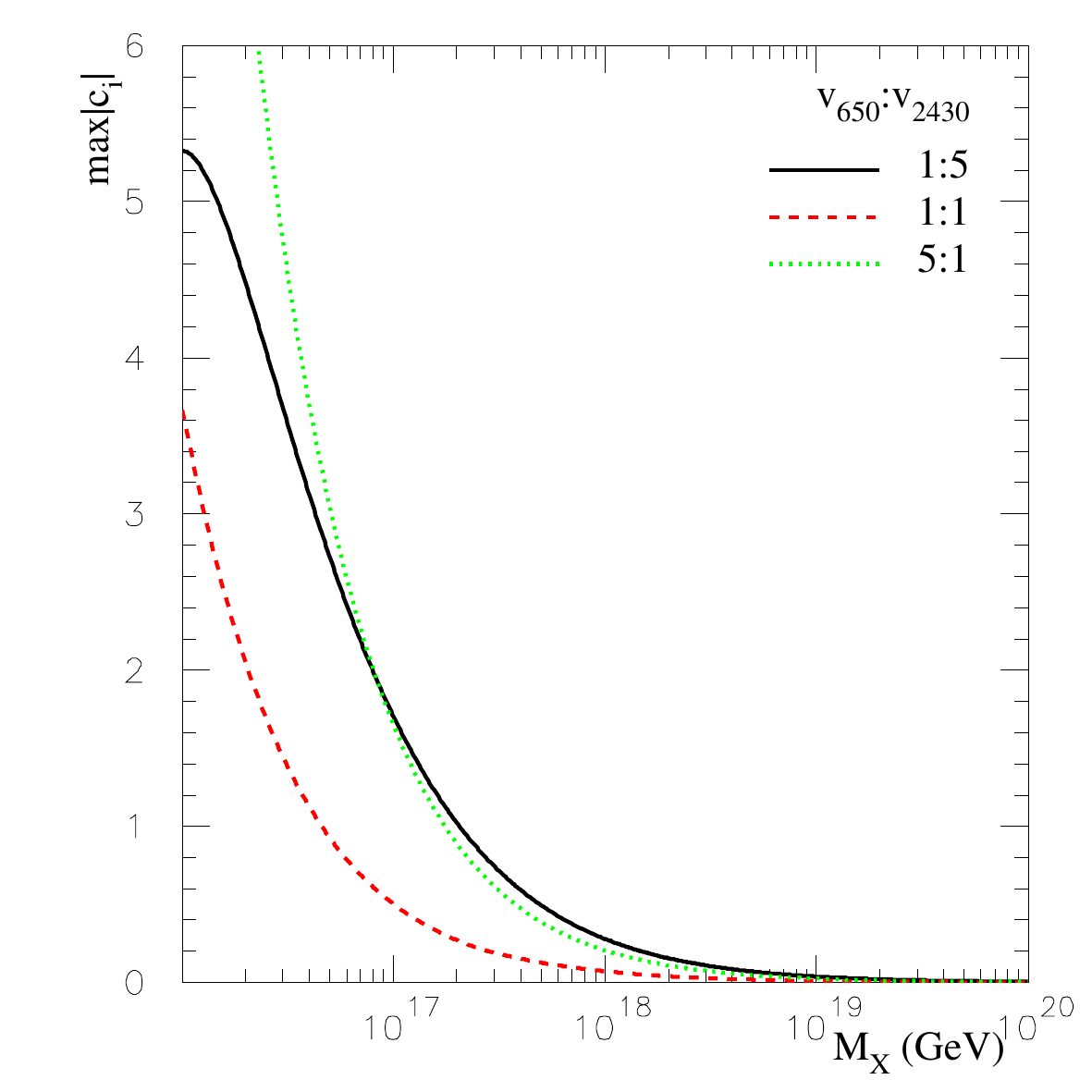}
\includegraphics[height=7cm,width=6cm]{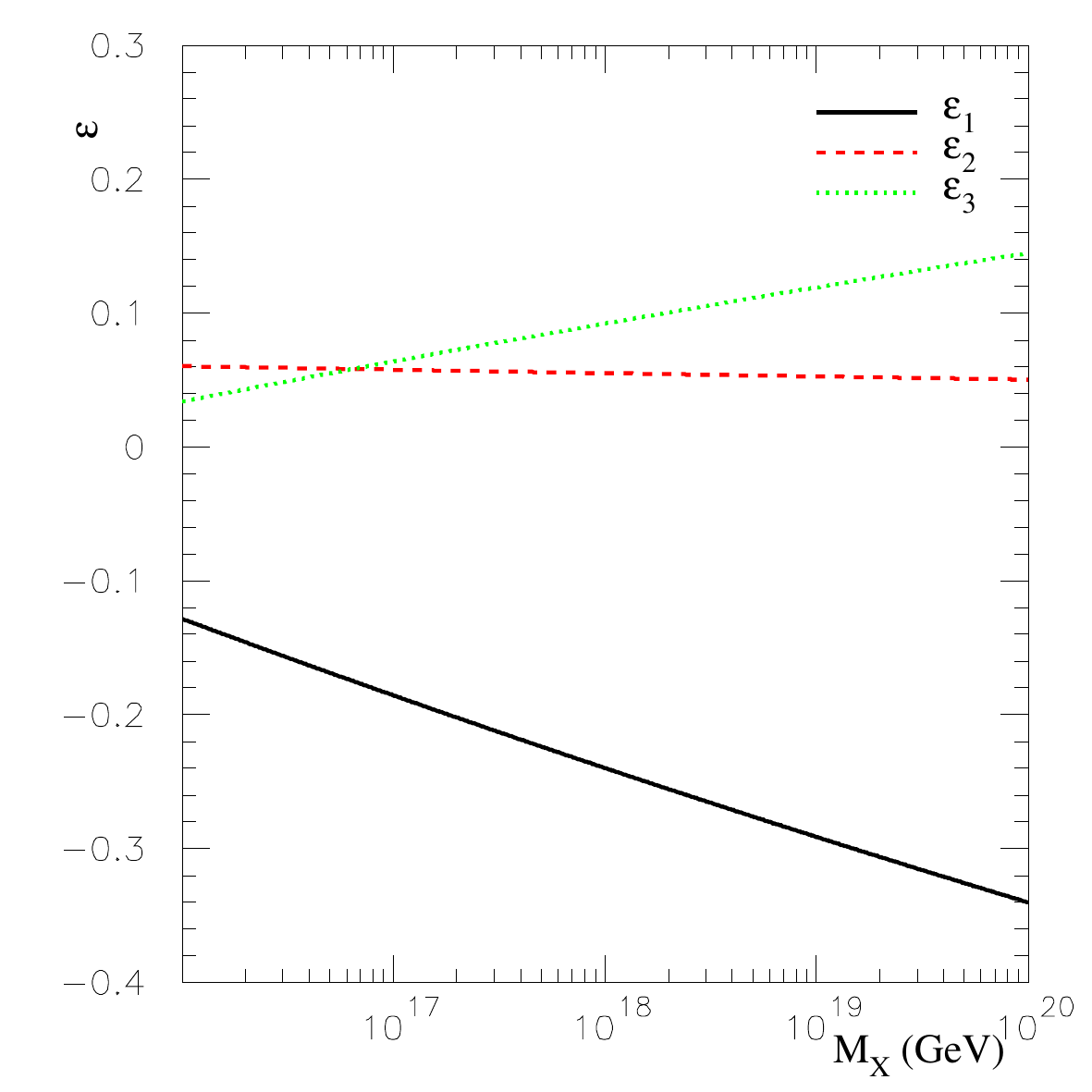}}
\caption{ \label{so10}
The maximal absolute value of Wilson coefficients as a function of
unification scale $M_X$ according to the transformation properties
under the subgroup embedding $SU(5)\subset SO(10) \subset E_6 $,
for two breaking chains, ${\bf 650} \to {\bf 54} \to {\bf 24}$ and
${\bf 2430} \to {\bf 210} \to {\bf 75}$.
The needed $\epsilon$s as a function of $M_X$ are plotted in the right.
The ratios of vacuum expectation values for the two breaking chains
are given as, $1:5$ (solid line), $1:1$ (dashed line) and $5:1$ (dotted line).
}
\end{figure}

\begin{figure}
\centerline{ \includegraphics[height=8cm,width=7cm]{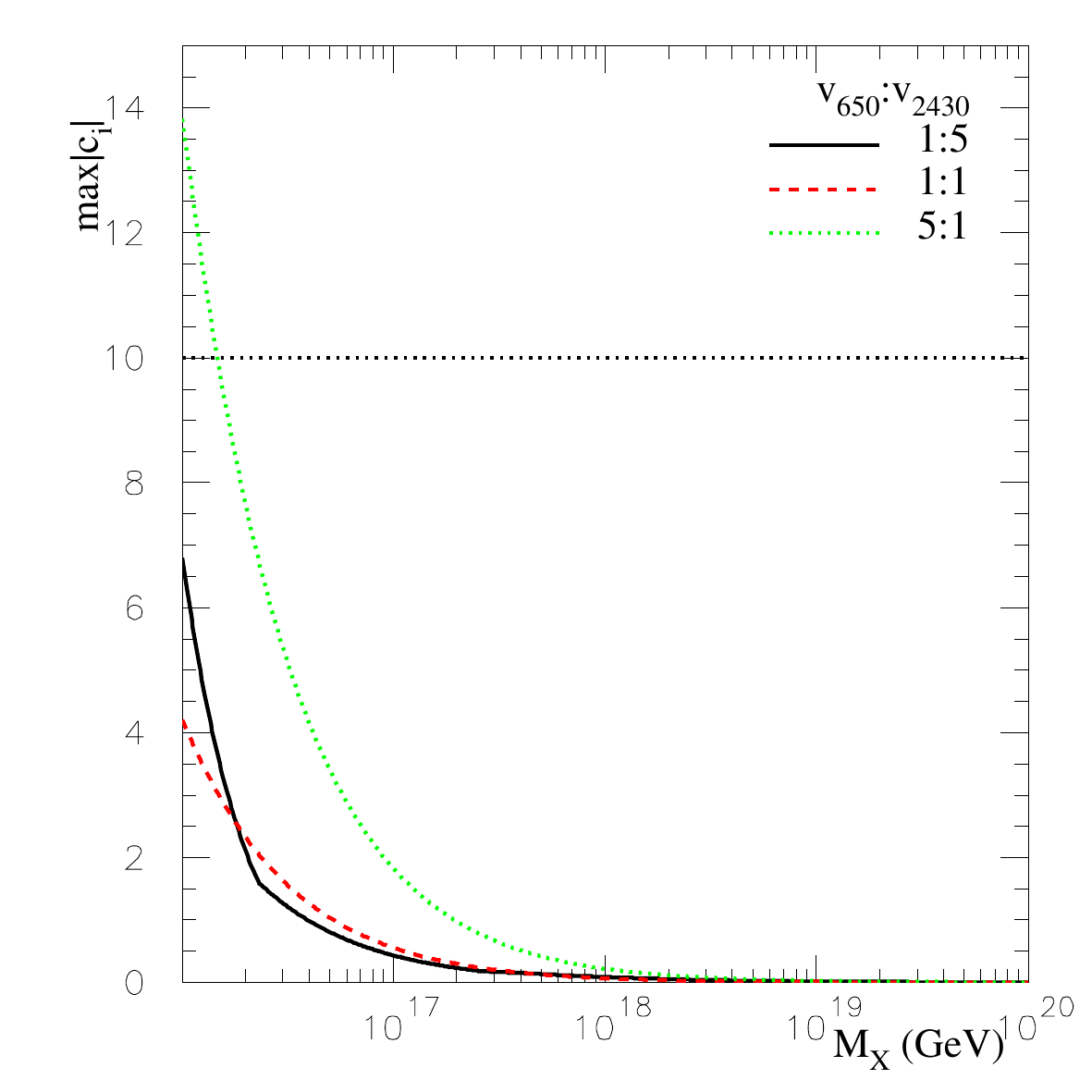} }
\caption{ \label{X}
The maximal absolute value of Wilson coefficients as a function of
unification scale $M_X$ according to the transformation properties
under the subgroup embedding $SU(3)_L \subset SU(6)\times SU(2)_X \subset E_6 $,
for two breaking chains, ${\bf 650} \to ({\bf 35, 1}) \to {\bf 1}$ and
${\bf 2430} \to ({\bf 405, 1}) \to {\bf 1}$.
The ratios of $\bf 650$ and $\bf 2430$ vacuum expectation values
are given as, $1:5$ (solid line), $1:1$ (dashed line) and $5:1$ (dotted line).
}
\end{figure}

\begin{figure}[htb]
\centerline{ \includegraphics[height=8cm,width=7cm]{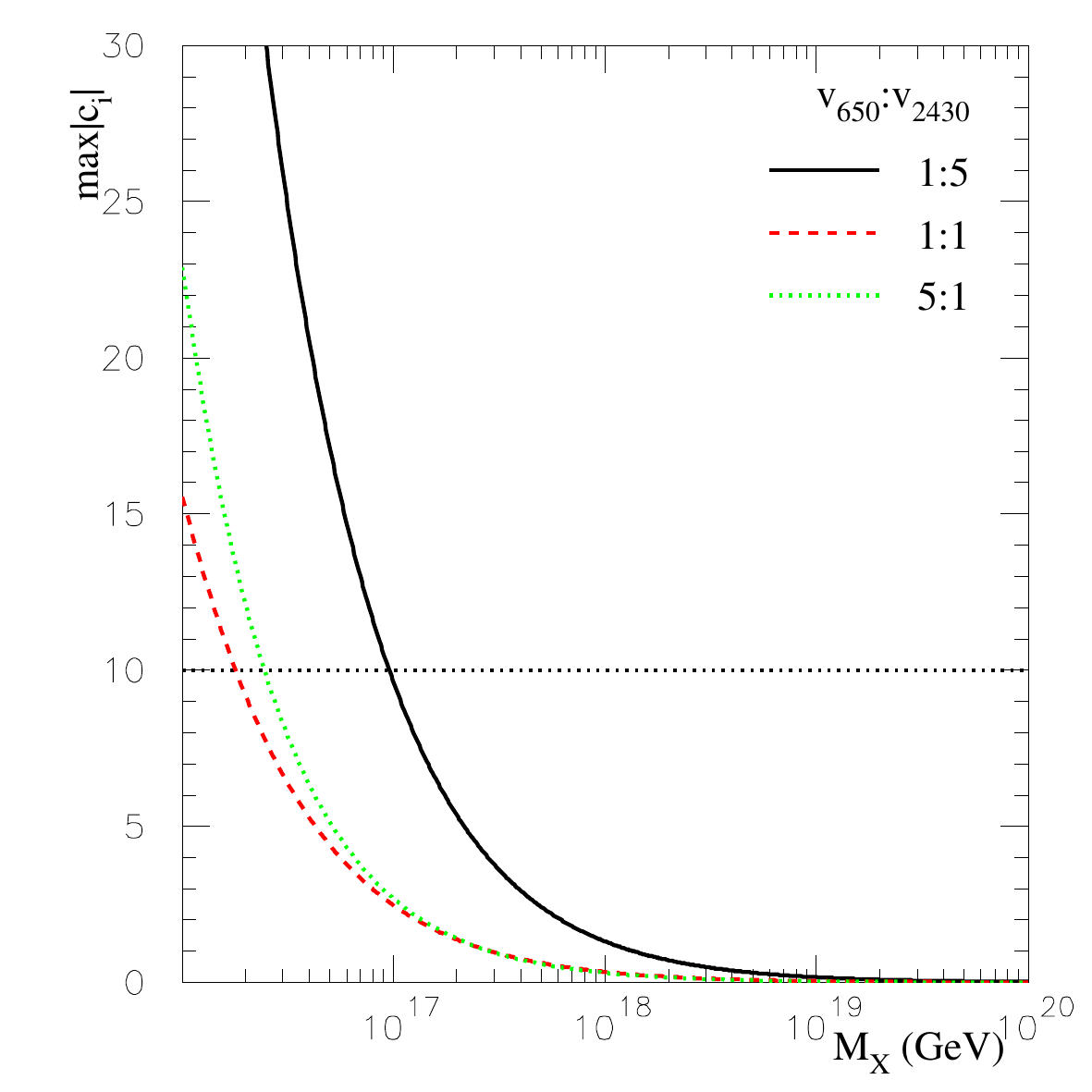} }
\caption{ \label{L}
The maximal absolute value of Wilson coefficients as a function of
unification scale $M_X$ according to the transformation properties
under the subgroup embedding $SU(3)_R \subset SU(6)\times SU(2)_L \subset E_6 $,
for two breaking chains, ${\bf 650} \to ({\bf 35, 1}) \to {\bf 1}$ and
${\bf 2430} \to ({\bf 1, 1}) \to {\bf 1}$.
The ratios of $\bf 650$ and $\bf 2430$ vacuum expectation values
are given as, $1:5$ (solid line), $1:1$ (dashed line) and $5:1$ (dotted line).
}
\end{figure}

\begin{figure}[htb]
\centerline{ \includegraphics[height=8cm,width=7cm]{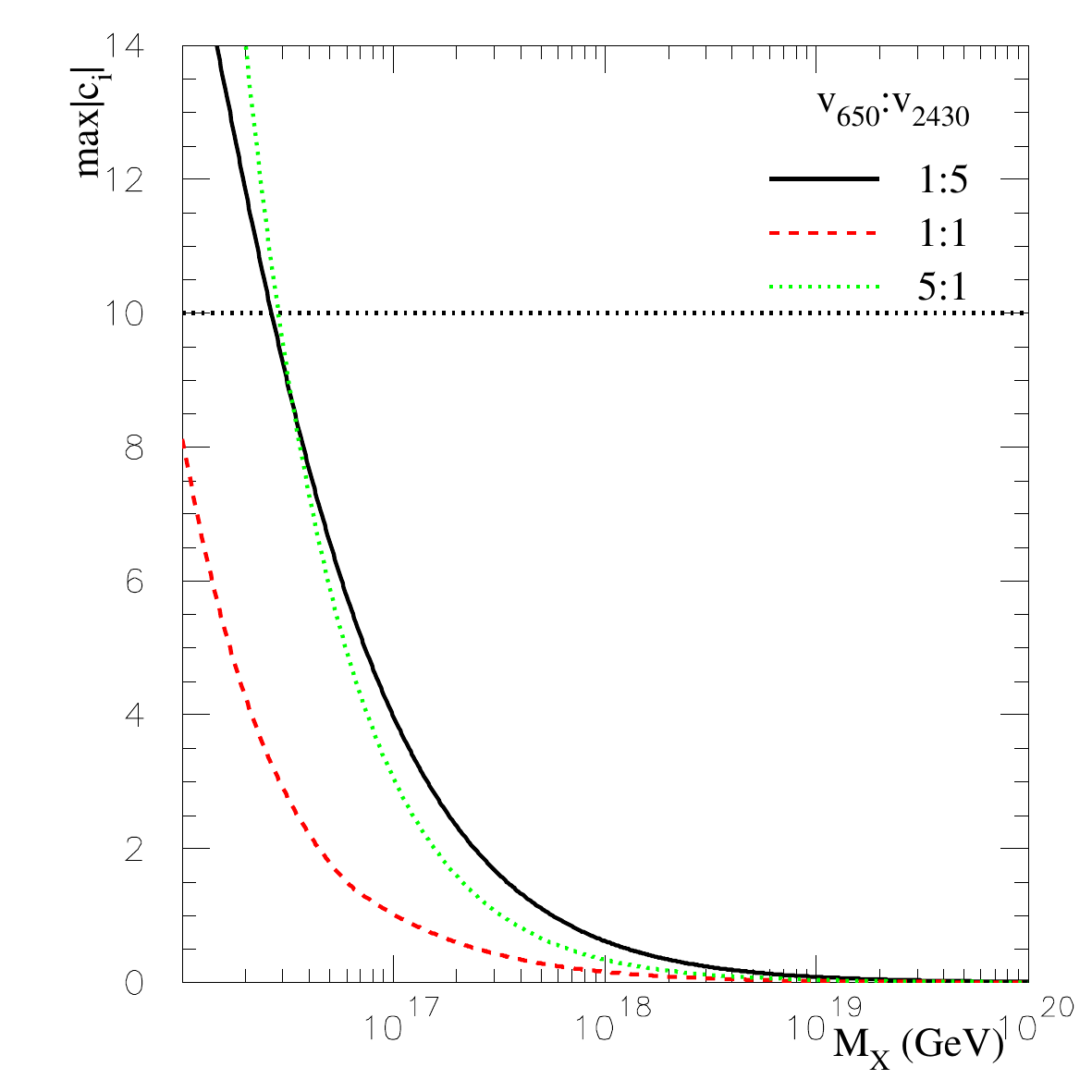} }
\caption{ \label{3}
The maximal absolute value of Wilson coefficients as a function of
unification scale $M_X$ according to the transformation properties
under the subgroup embedding $SU(3)_L \times SU(3)_R \subset E_6 $,
for two breaking chains, ${\bf 650} \to ({\bf 1, 1})_2$ and
${\bf 2430} \to ({\bf 8, 1})$.
The ratios of $\bf 650$ and $\bf 2430$ vacuum expectation values
are given as, $1:5$ (solid line), $1:1$ (dashed line) and $5:1$ (dotted line).
}
\end{figure}

In the following, we study the unification of gauge couplings
in the case of two dimension-5 operators
with two different Higgs multiplets
from $\bf 650$ and $\bf 2430$ of $E_6$ respectively.
In order to achieve continuously varied unification scale $M_X$,
we have four variables, two VEVs of Higgs multiplets,
$v_{\bf 650}$ and $v_{\bf 2430}$, and two Wilson coefficients.
Without fine-tuning for the VEVs,
we fix the ratio of $\bf 650$ and $\bf 2430$
in three cases, $1:5$, $1:1$ and $5:1$, in our figures for an illustration.
Also, for the VEVs of the Higgs multiplets,
we assume they account for half of
the average gauge boson squared mass~\cite{Calmet:2009hp}. % as before.
Then, we are left with two Wilson coefficients for the unification.
The maximal absolute values of Wilson coefficients as a function of
unification scale $M_X$ are given in Fig.~\ref{so10},\ref{X},\ref{L},\ref{3}
for different embedding of subgroup into $E_6$.

%\clearpage

%\clearpage
The numerical results of the subgroup embedding
$SU(5)\subset SO(10)\times U(1) \subset E_6 $ are shown in Fig.~\ref{so10}.
We choose two breaking chains, ${\bf 650} \to {\bf 54} \to {\bf 24}$ and
${\bf 2430} \to {\bf 210} \to {\bf 75}$,
as an example for the study.
For all three cases of the ratios of Higgs VEVs,
we have Wilson coefficients less than $10$,
max$|c_i| \le 10$, while unification scale $M_X > 10^{16}$GeV.
The needed $\epsilon$s for the unification are also given in Fig.~\ref{so10},
which are independent of the ratios of Higgs VEVs.
The $\epsilon_2$ is varied smoothly with the scale $M_X$,
but the needed $\epsilon_1$ is negative and larger absolute value is required.
Similar results for the other embeddings,
$SU(3)_L \subset SU(6)\times SU(2)_X \subset E_6 $,
$SU(3)_R \subset SU(6)\times SU(2)_L \subset E_6 $ and
$SU(3)_L \times SU(3)_R \subset E_6 $,
are shown in Fig.~\ref{X},\ref{L},\ref{3}, respectively.
In order to obtain %{\color{red}satisfy }
max$|c_i| \le 10$, larger unification scale
$M_X$ are needed for different ratios of Higgs VEVs.
A comment is that for different subgroup embedding of $E_6$,
we may have different sets of values of $\epsilon$s
at a given unification scale $M_X$.

\begin{figure}[ht]
\centerline{ \includegraphics[height=8cm,width=7cm]{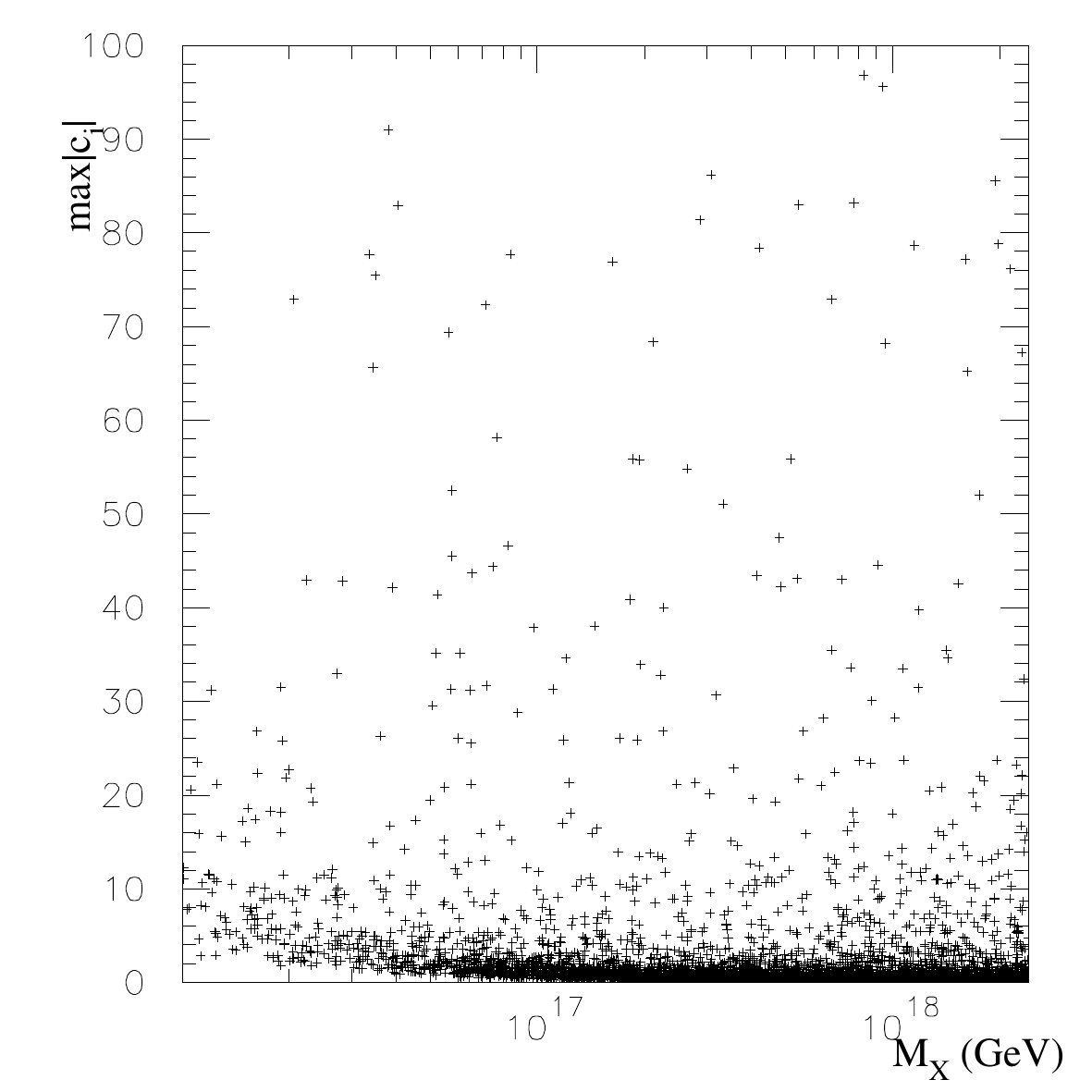} }
\caption{ \label{so10_scan}
The maximal absolute value of Wilson coefficients as a function of
unification scale $M_X$
according to the transformation properties
under the subgroup embedding $SU(5)\subset SO(10) \subset E_6$,
for all breaking chains of ${\bf 650}$ and ${\bf 2430}$
with random VEVs and Wilson coefficients.
}
\end{figure}

\begin{figure}[h]
\centerline{ \includegraphics[height=8cm,width=7cm]{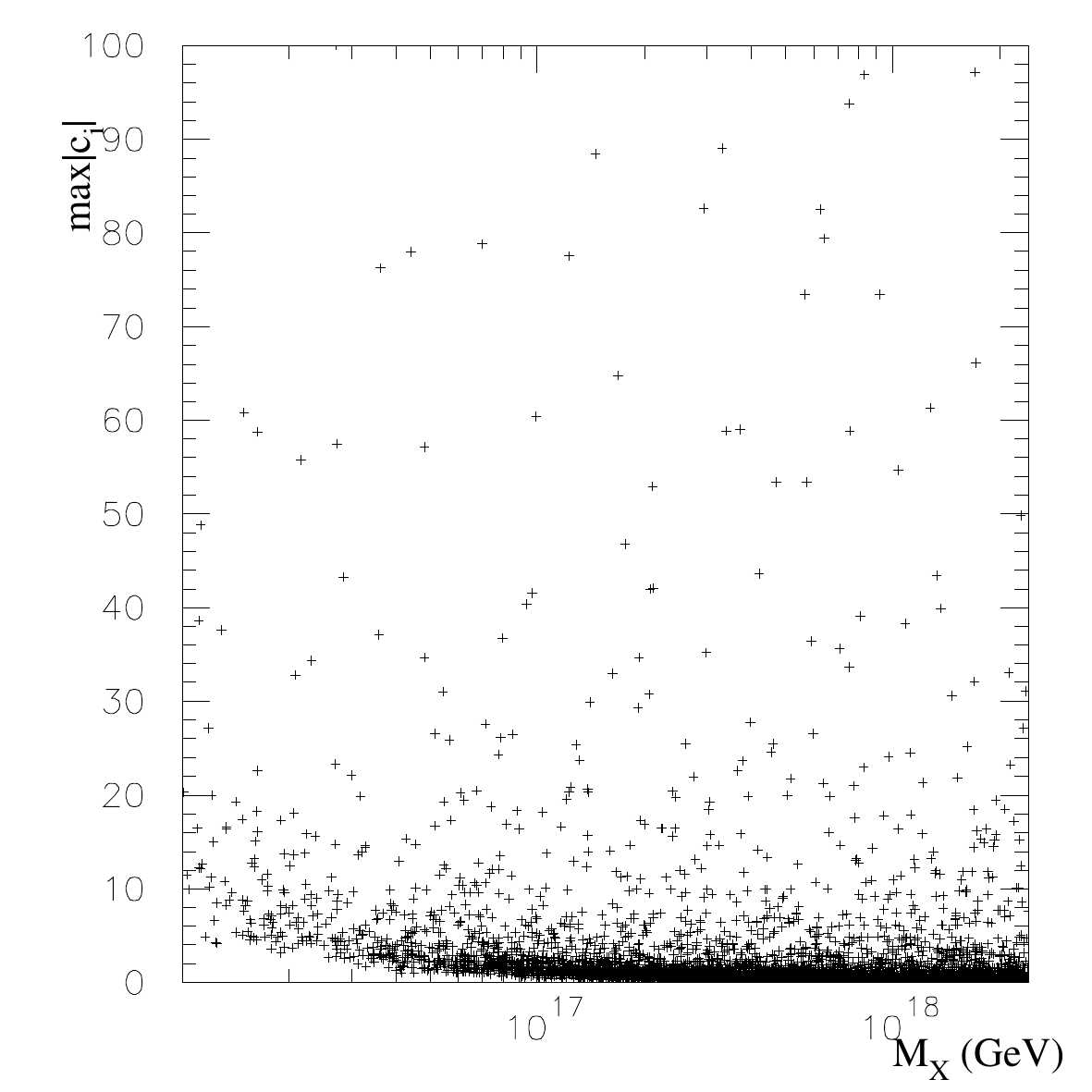} }
\caption{ \label{X_scan}
The maximal absolute value of Wilson coefficients as a function of
unification scale $M_X$ of
according to the transformation properties
under the subgroup embedding
$SU(3)_R \subset SU(6) \times SU(2)_X \subset E_6 $,
for all breaking chains of ${\bf 650}$ and ${\bf 2430}$
with random VEVs and Wilson coefficients.
}
\end{figure}

In the case of two dimension-5 operators %$\bf 650$ and $\bf 2430$
with more than two breaking chains,
we scan the maximal absolute value of Wilson coefficients as a function of
unification scale $M_X$ for
subgroup embeddings $SU(5)\subset SO(10) \subset E_6$ and
$SU(3)_L \subset SU(6) \times SU(2)_X \subset E_6 $
with random VEVs of different breaking chains
and random ratios of Wilson coefficients of the two operators
in Fig.~\ref{so10_scan} and Fig.~\ref{X_scan}.
For subgroup embedding $SU(5)\subset SO(10) \subset E_6$,
there are actually four different ratios among $v_{s,z}^{k}$,
and for subgroup embedding $SU(3)_L \subset SU(6) \times SU(3)_X \subset E_6 $,
we have eight different ratios.
As a result, the unification are much easier for the latter.
Most of the points in both cases have satisfed the successful unification
with max$|c_k| \le 10$.

\clearpage

\clearpage

As a summary of this section,
we conclude that nonsupersymmetric models of $E_6$ grand unification
can be obtained through effects of dimension-5 operators.
Comparing with the other groups like $SU(5)$ and $SO(10)$,
it is much easier to achieve the successful unification
with natural Wilson coefficient $c_k$ and
continuously varied unification scale $M_X$.
Thus including effects of quantum gravity provides
a greater probability for building a realistic nonsupersymmetric model
in $E_6$ GUT.

\clearpage
\section{The constraint from the proton decay}

As it is well-known, the key predictions in GUT,
despite of the details of model-building,
are the gauge coupling unification and the proton decay.
It has been pointed out~\cite{Tobe:2003yj,Nath:2006ut} that
contributions to the proton decay mediated by superheavy Higgs,
which are mainly due to color triplets and very model-dependent,
are less important than those mediated by superheavy gauge bosons.
Therefore, we limit ourselves to discuss only the contributions from
the gauge dimension-6 operators.
The grand unification breaking will bring in heavy thresholds.
Assuming one-step breaking of the grand unified gauge group $E_6$
to the SM $G_{321}$ at the unification scale $M_{X}$ for simplicity,
the average squared mass of the non-$G_{321}$ singlet gauge bosons
(usually called "superheavy" gauge bosons) is given by,
\be
\overline{m^2}_{ab}=\sum_i \frac{C(r_i)}{66}g_G^2v_i^2\,\delta_{ab}\equiv M_{\rm HG~ b}^{2}\,\delta_{ab}~~~\text{for}~a,b=13,\,\ldots,78~,
\ee
where the sum runs over all Higgs multiplets $i$
that acquire nonzero vacuum expectation value $v_i$ at $M_X$ ,
i.e., in addition to the non-$G_{321}$ singlet Higgs contained
in Eq. (\ref{dim5}), all other Higgs multiplets
which are necessary to realize the gauge symmetry breaking chain
in a specific model.
In order to guarantee correct use of the running gauge couplings
in the SM (see the previous section),
it is necessary to require $M_{HG}$ close to
the grand unified symmetry breaking scale,
i.e., the unification scale $M_X$, in the case of one-step breaking.
For the purpose of definiteness and
omitting heavy threshold effects, we take $M_{HG}=M_X$.
Then, the proton lifetime due to superheavy gauge boson exchange
can be written as~\cite{Tobe:2003yj,Nath:2006ut,Dorsner:2004xa},
\be \label{thpd}
\tau_p=C M_X^4 /(\alpha_G^2 m_p^5),
\ee
where C is a coefficient containing all information about
the flavor structure of the theory and $m_p$ is the mass of proton.
The newest experimental bound on the proton lifetime
for the channel $p\rightarrow e^+\pi^0$ is~\cite{TheSuper-Kamiokande:2017tit},
\be \label{expd}
\tau_{p\rightarrow e^+\pi^0} > 1.6 \times 10^{34}~{\rm years}.
\ee
Combining Eq. (\ref{thpd}) and Eq. (\ref{expd}), one has
\be \label{mxl}
M_X> (40 \alpha_G)^{1/2}(1/C)^{1/4} 4.3\times 10^{15} GeV.
%10^{1/4}=1.7784
\ee
C is order of 1, $C \sim O(1)$.
$\alpha_G=1/70\sim 1/40$ (see Table II and Figures above).
Even when C is as small as 0.1, the constraint from the proton decay is just
$M_X > 8.9\times 10^{15} {\rm GeV}$.

We can see from the Table II and figures in the previous section
that most of the cases can easily satisfy the %constraint {\color{red}of
bound Eq. (\ref{mxl}).
In the case of only one breaking chain,
${\bf 2430} \to ({\bf 405, 1}) \to {\bf 1}$
and ${\bf 2430} \to ({\bf 189, 1}) \to {\bf 1}$,
have the unification scale $6.4\times 10^{16}GeV$
and $3.6\times 10^{18}GeV$, respectively.
Both lie well above the bound Eq. (\ref{mxl}).
For the case of ${\bf 2430} \to {\bf 210} \to {\bf 75}$,
the unification scale is $3.7\times 10^{15}GeV$,
which is very close to the bound  Eq. (\ref{mxl})
and can be easily adjusted (say, to increase the value of v)
to satisfy the bound.
With only one dimension-5 operator,
using two (or more) different breaking chains,
one can achieve successful gauge coupling unification
with continuously varied unification scale $M_X$,
as long as the ratio of two vevs varies with $M_X$.
Thus, the bound Eq. (\ref{mxl}) is satisfied.
Looking at the results shown in the last section,
the same remains in the case of two dimension-5 operators
with two different Higgs multiplets from $\bf 650$ and $\bf 2430$
of $E_6$ respectively.

\section{Summary and discussion}

It has been pointed out that gauge coupling unification condition
is modified due to the effects of dimension-5 operators.
We have investigated the gauge coupling unification in $E_{6}$
without SUSY under modified gauge coupling unification condition.
For this purpose, considering several maximal subgroups
$H=SO(10)\times U(1)$, $H=SU(3)\times SU(3)\times SU(3)$,
$H=SU(2)\times SU(6)$ of $E_{6}$
and the usual breaking chains for a specific maximal subgroup,
we have derived and given all of the Clebsch-Gordan coefficients
$\Phi^{(r)}_{s,z}$ associated with $E_6$ breaking to the SM.
We have also presented the structure constants of $E_6$ in the usual form,
which are mostly used by physicists and in the study of GUT.
Results on the gauge coupling unification
show that, for a single dimension-5 operator,
realizing the gauge coupling unification
under modified gauge coupling unification condition in $E_6$ GUT
is easier than that in $SO(10)$ GUT,
since there are more maximal subgroups,
and, for a specific maximal subgroup,
there are more breaking chains,
so that one has much freedom to choose ratios
among  the nonzero vacuum expectation values $v_{s,z}^{k}$
of the Higgs multiplets $H_k$.
We have also analyzed the constraint on the unification scale $M_{X}$
from the newest data of the proton decay.
It is shown that most of cases studied
in the section IV satisfy the constraint easily. % (\ref{mxl})

In the effective field theory spirit,
operators of dimension higher than 5 are also present,
e.g., a dimension-6 operator generalization of Eq. (\ref{dim5}),
\begin{equation}\label{dim6operators}
{\cal L}=\frac{c_6}{M_{Pl}^2}H_1H_2G_{\mu\nu}G^{\mu\nu}+\ldots~.
\end{equation}
After the Higgs multiplets acquire vevs at the scale $M_X=M_{Pl}/O(1)$,
they can contribute to the gauge kinetic terms as well,
\begin{equation} \label{expand}
{\cal L}=\sum_{i=1}^3-\frac{1}{4}\left(1+\epsilon_i+\epsilon_i^{(6)}+\ldots\right)G_{(i)\mu\nu}^aG_{(i)}^{a\mu\nu}~,
\end{equation}
where the corrections (see Eq. (\ref{epsilond})), %$\epsilon_i\sim c_k\phi_i^k M_X/g_GM_{Pl}$
\begin{eqnarray} %\label{epsilon}
\epsilon_i =\frac{c_k}{M_{Pl}} <H_k>
= \sum_k \frac{c_k}{M_{Pl}} \sum_{s,z} v_{s,z}^k \phi_i^{k~sz}, ~~~i = 1, 2, 3, ~~~~ \nonumber \\
 \phi_1^{k~sz}\equiv -h3^{k}_{s,z},~~\phi_2^{k~sz}\equiv -h2^{k}_{s,z},~~\phi_3^{k~sz}\equiv -h1^{k}_{s,z}, \nonumber
\end{eqnarray}
from the dimension-5 operators (\ref{dim5}) and
the corrections $\epsilon_i^{(6)} $ from
the dimension-6 operators (\ref{dim6operators}).
We can estimate the size of $\epsilon_i^{(6)}$.
For example, taking $H_1={\bf 27}$ and $H_2={\bf \bar{27}}$, we have,
\be\label{cg} {\bf 27}\times{\bf \bar{27} }= {\bf 1}+{\bf 78}+{\bf 650}.
\ee
So we can use $cg_k H_k v$,
where $v\sim \frac{M_X}{g_G}$ is the average vev of the $H_1,H_2$
and $cg_k$ ($k=1,2,3$ corresponding to the representation
${\bf 1},{\bf 78},{\bf 650}$ respectively)
is the Clebsch-Gordan coefficient from the decomposition of
the Kronecker product ${\bf 27}\times{\bf \bar{27} }$ (see (\ref{cg})),
instead of $H_1H_2$, in (\ref{dim6operators}).
For specific, set $k=3$, i.e., the ${\bf 650}$, then, we have,
\be \label{d6de}
{\cal L}=\frac{c_6}{M_{Pl}^2} \frac{M_X}{g_G} cg_3 H_3 G_{\mu\nu}G^{\mu\nu}+\ldots~,
\ee
which leads to $\epsilon_i^{(6)}=\frac{c_6 M_X}{g_G M_{Pl}^2} cg_3 <H_3>$.
For $k=3$, $\epsilon_i =\frac{c_3}{M_{Pl}} <H_3>$.
Therefore,
$\epsilon_i^{(6)}\sim \frac{c_6 cg_3 M_X}{c_3 g_G M_{Pl}} \epsilon_i  $.
With $c_6\sim c_3$, $cg_3\leq 1$ and $M_X=M_{Pl}/O(1)$,
effects of dimension-6 operators might be the same order
of those of dimension-5 operators. Therefore,
%Depending on the group theory factors $\delta^{(6)}_s$ (analogous to the %$\delta_s$ in Table \ref{deltaSU5table}) and on the constant in the relation %$M_X=M_{Pl}/O(1)$,
the expansion of Eq. (\ref{expand}) might not
or might be controlled perturbatively.
If it is not, one cannot claim perturbative gauge-gravity unification
at the Planck scale.
However, the fact that the modification
of the gauge coupling unification condition of Eq. (\ref{gaucon}) allows us
in principle to adjust the unification scale to
a higher scale $\sim M_{Pl}$
could at least be taken as a hint that
gauge-gravity unification is a possible scenario,
even if the necessary parameter values or
the last piece of the evolution cannot be computed perturbatively.

There is an interesting subject in the effective theory framework.
That is, the following operators of dimension-5 are also probably present,
\be \label{axion}
{\cal L} =\frac{c_{k}}{M_{Pl}} G^{a}_{\mu\nu} \tilde{G}^{b \mu\nu} \tilde{H}_{k}^{ab},
\ee
where $a,b$ are group indices, $k$ labels different Higgs multiplets
and $\tilde{G}^{b \mu\nu}$ is dual of gauge field strength $G^b_{\mu\nu}$.
If one assumes that when $E_6$ breaks into $G_{321}$,
one of the other two U(1)'s ($U_{V}(1), U_{V'}(1)$) is anomalous
Peccei-Quinn $U(1)$.
After spontaneously breaking of anomalous Peccei-Quinn $U(1)$ symmetry,
the associated pseudo Goldstone boson, axion,
is to become a component of the corresponding vector boson
(say, $Z_V$), since now this $U(1)$ is a local symmetry.
When $\tilde{H}_k$ acquires the nonzero vacuum expectation values,
from Eq.(\ref{axion})
and  the relevant Clebsch-Gordan coefficients $\Phi^{(r)}_{s,z}$,
one can estimate the size of the coupling of $Z_V$.
It might be instructive to study the relation of this term
to strong CP and axion physics.

It is straightforward to generalize the study
of the gauge coupling unification in $E_{6}$
including the effects of dimension-5 operators to SUSY GUT.
In this case, gaugino mass ratios can be read %off
from the Tables shown in Appendix A, since gauginos belong to
the same multiplets as gauge bosons in SUSY.
That is, for the flipped embedding,
\be M_3:M_2:M_1= h1^{k}_{s,z}:h2^{k}_{s,z}:h3'^{k}_{s,z},
\ee
where $M_a$ are the gaugino masses
and $a=3,2,1$ corresponds the $SU(3)$, $SU(2)$, $U(1)$ of the SM,
i.e., the gluino, wino and bino masses.
The results agree with those in corresponding Tables
given by S.P. Martin~\cite{Martin:2009ad}.

For model building of $E_6$ GUT with effects of dimension-5 operators,
there are several important problems,
such as the doublet-triplet splitting, neutrino mass hierarchy, etc.,
which need to be answered.
However, this is beyond the scope of this paper.
One should study them in the future.

%\clearpage

\section{Appendix A: Clebsch-Gordan coefficients
associated with $E_6$ breaking to the SM}

All of the Clebsch-Gordan coefficients $\Phi^{(r)}_{s,z}$
associated with $E_6$ breaking to the SM,
in different bases $\{s,z\}$,
up to a uniform normalization constant
for different representations $r$
have been derived and results are given in this Appendix.
For the subgroup $H=SO(10)\times U(1)$,
the analysis and results have been given in the Ref. \cite{Huang:2014zba}.
We start from $H=SU(6)\times SU(2)$.

%\begin{sidewaystable}[t]
%\newsavebox{\tablebox}
%\begin{lrbox}{\tablebox}
%\clearpage

\begin{table}[ht] %\scriptsize
%\tiny
%\begin{center}
\tabcolsep=0.01cm
\footnotesize{
\begin{tabular}{c|c|c||c|c|c|c|c|c|c|c|c|c|c|c|c|c|c|c}
\hline
$E_6$ & SU(6)$\times$ & $SU(3)_L$ z &
 $h1^{r}_{s,z}$ & $h2^{r}_{s,z}$ & $h3^{r}_{s,z}$ & $h4^{r}_{s,z}$ &
 $h5^{r}_{s,z}$ & $h6^{r}_{s,z}$ & $h7^{r}_{s,z}$ & $h8^{r}_{s,z}$ &
 $h9^{r}_{s,z}$ & $h10^{r}_{s,z}$ & $h11^{r}_{s,z}$ & $h12^{r}_{s,z}$ &
 $h13^{r}_{s,z}$ & $h14^{r}_{s,z}$ & $h15^{r}_{s,z}$ & $N^r_{s,z}$  \\
 & SU(2)$_X$ s & & & & & & & & & & & & & & & & \\
\hline
\hline
${\bf 1}$ & ${\bf 1}$ 1 & ${\bf 1}$  1 &  1 & 1 & 1 & 1 & 1&  1& 1& 1 & 1 & 1 & 1 & 1&  1& 1& 1 &$\frac{1}{\sqrt{78}}$ \\
\hline
\hline
${\bf 650}$ & $({\bf 1}, {\bf 1})$ 1& ${\bf 1}$  1 &  $1$ & $1$ &
 $1$ & -$\frac{1}{2}$ & -$\frac{11}{4}$ & -$\frac{1}{2}$ &
 $1$ & -$5$ & -$\frac{5}{4}$ & $1$ & -$\frac{1}{2}$ &
 -$\frac{1}{2}$ & -$\frac{1}{2}$ & $1$ & -$\frac{1}{2}$ &
 $\frac{1}{2\sqrt{30}}$ \\
% \hline
  & $({\bf 35}, {\bf 1})$ 2& ${\bf 1}$ 1 & $1$ & -$1$ & -$\frac{1}{5}$ & $0$ &
 -$\frac{3}{10}$ &  $0$ & $0$ & $0$ & -$\frac{1}{2}$ & $0$ & $0$ &
 $0$ & $0$ & -$1$ &
 $0$ & $\frac{1}{4}$ \\
%\hline
  &     & ${\bf 8}$ 2 & $0$ & $1$ & $\frac{3}{5}$ & $0$ & -$\frac{3}{5}$ &
 $0$ & -$1$ &
 $0$ & -$1$ & $\frac{1}{2}$ & $0$ & $0$ & $0$ & -$\frac{1}{2}$ &
 $0$ & $\sqrt{\frac{2}{41}}$ \\
%\hline
       & $({\bf 189}, {\bf 1})$ 3  & ${\bf 1}$ 1 & $1$ & $1$ & -$3$ &
 $0$ & $\frac{1}{2}$ & $0$ & -$\frac{2}{3}$ & $0$ & -$\frac{1}{2}$ &
 -$\frac{2}{3}$ & $0$ &
 $0$ & $0$ & $1$ & $0$ & $\frac{1}{2}\sqrt{\frac{3}{34}}$ \\
%\hline
  &     & ${\bf 8}$ 2 & $0$ & $1$ & -$1$ & $0$ & -$\frac{3}{2}$ &
 $0$ & $\frac{1}{3}$ &
 $0$ & $\frac{3}{2}$ & -$\frac{1}{6}$ & $0$ & $0$ & $0$ & -$\frac{1}{2}$ &
 $0$ & $\sqrt{\frac{6}{83}}$ \\
%\hline
\hline
\hline
${\bf 2430}$ & $({\bf 1}, {\bf 1})$ 1& ${\bf 1}$ 1 &  $1$ & $1$ &
 $1$ & -$\frac{7}{6}$ & $\frac{101}{36}$ & -$\frac{7}{6}$ & $1$ &
 $\frac{35}{9}$ & $\frac{25}{12}$ & $1$ &
 -$\frac{7}{6}$ & -$\frac{7}{6}$ & -$\frac{7}{6}$ & $1$ & -$\frac{7}{6}$ &
 $\frac{3}{2}\sqrt{\frac{3}{910}}$ \\
%\hline
 & $({\bf 189}, {\bf 1})$ 2  & ${\bf 1}$ 1 &  $1$ & $1$ & -$3$ & $0$ &
 -$\frac{15}{4}$ & $0$ & -$\frac{2}{3}$ & $0$ & $\frac{15}{4}$ &
 -$\frac{2}{3}$ & $0$ & $0$ & $0$ & $1$ &
 $0$ & $\frac{1}{2\sqrt{85}}$ \\
%%\hline
        &   & ${\bf 8}$ 2 & $0$ & $1$ & -$1$ & $0$ & $\frac{9}{4}$ &
 $0$ & $\frac{1}{3}$ &
 $0$ & -$\frac{9}{4}$ & -$\frac{1}{6}$ & $0$ & $0$ & $0$ & -$\frac{1}{2}$ &
 $0$ & $\sqrt{\frac{2}{41}}$ \\
%\hline
 & $({\bf 405}, {\bf 1})$ 3& ${\bf 1}$ 1 & $1$ & $1$ & $\frac{33}{5}$ & $0$ &
 $\frac{9}{10}$ & $0$ & -$\frac{4}{3}$ & $0$ & $\frac{3}{2}$ &
 -$\frac{4}{3}$ & $0$ & $0$ & $0$ &
 $1$ & $0$ & $\frac{1}{4\sqrt{7}}$ \\
%\hline
 &                         & ${\bf 8}$ 2 & $0$ & $1$ & $\frac{19}{5}$ & $0$ &
 -$\frac{9}{5}$ & $0$ & $\frac{5}{3}$ & $0$ & -$3$
 & -$\frac{5}{6}$ & $0$ & $0$ & $0$ &
 -$\frac{1}{2}$ & $0$ & $\sqrt{\frac{2}{185}}$ \\
%\hline
 &                        & ${\bf 27}$ 3 & $0$ & $1$ & $\frac{9}{5}$ & $0$ &
 $\frac{27}{10}$ & $0$ & $0$ & $0$ & $\frac{9}{2}$ & $0$ & $0$ & $0$ & $0$ &
 -$3$ & $0$ & $\frac{1}{2\sqrt{30}}$ \\
\hline
\end{tabular}
}
%\end{lrbox}
%\rotatebox[]{0}{\usebox{\tablebox}}
%\end{center}
\caption{
The diagonal part, $hj$ $(j=1,2,...,15)$ in Eq. (\ref{phir})
of the SM singlets $\Phi_{s,z}^r$ in each of the irreps $r$
according to their transformation properties
under the $SU(3)_L \subset SU(6)\times SU(2)_X \subset E_6 $ subgroup.}
%\caption{The diagonal part in Eq. (\ref{phir}) of the standard model singlets $\Phi_{s,z}^r$ in each of the irreps $r$ (see Eq. (\ref{reps})) of $E_6$, classified according to their transformation properties under the $SU(5)\subset SO(10) \subset E_6 $ subgroup (SO(10):second column; SU(5): third column  ), in the explicit version with the conventions described in the section II of the text, for the normal embedding. They agree between the normal and the flipped embedding of $G_{321}\subset E_6$, except for the $h3_{s,z}^r$, $h5_{s,z}^r$, $h9_{s,z}^r$. The entries $h3_{s,z}^{\prime r}$, $h5_{s,z}^{\prime r}$, $h9_{s,z}^{\prime r}$ for the flipped embedding are listed in Table III. $N^r_{s,z}$ is the normalization constant which makes $Tr(\Phi^{r}_{s,z} \Phi^{r}_{s,z})=1$ for each irrep r with specific s and z. (For both the normal and the flipped embedding, the $SU(5)$ here contains $SU(3)_C\times SU(2)_L$.)\label{su5}}
\end{table}
%\end{sidewaystable}

\begin{table}[b]%\footnotesize
%\begin{center}
\tabcolsep=0.1cm
\footnotesize{
\begin{tabular}{c||c|c||c|c|c||c|c|c|c|c|c|c|c|c|c|c|c|c}
\hline
$E_6$ & SU(6)$\times$ & SU(3)$_L$ z & $h35^{r}_{s,z}$ & $h39^{r}_{s,z}$ &
 $h59^{r}_{s,z}$ & $h3^{\prime r}_{s,z}$ & $h5^{\prime r}_{s,z}$ &
 $h9^{\prime r}_{s,z}$ & $h35^{\prime r}_{s,z}$ &
 $h39^{\prime r}_{s,z}$ & $h59^{\prime r}_{s,z}$  \\
 & SU(2)$_X$ s & & & & & & & & & & \\
\hline
\hline
${\bf 1}$ & ${\bf 1}$ 1 & ${\bf 1}$  1 & 0& 0&0 &0& 0&0 &0& 0&0  \\
\hline
\hline
${\bf 650}$ & $({\bf 1}, {\bf 1})$ 1& ${\bf 1}$  1 & $0$ &
 $0$ & -$\frac{3}{4}\sqrt{15}$ & $1$ & -$\frac{11}{4}$ & -$\frac{5}{4}$ &
 $0$ & $0$ & -$\frac{3\sqrt{15}}{4}$ \\
%\hline
  & $({\bf 35}, {\bf 1})$ 2& ${\bf 1}$ 1 & $\frac{\sqrt{3/2}}{5}$ &
 -$\frac{1}{\sqrt{10}}$ & $\frac{\sqrt{3/5}}{2}$ & -$\frac{1}{5}$ &
 -$\frac{3}{10}$ & -$\frac{1}{2}$ &
 $\frac{\sqrt{3/2}}{5}$ & -$\frac{1}{\sqrt{10}}$ & $\frac{\sqrt{3/5}}{2}$  \\
%\hline
  &     & ${\bf 8}$ 2 & -$\frac{\sqrt{3/2}}{10}$ & $\frac{1}{2\sqrt{10}}$ &
 $\sqrt{3/5}$ & $\frac{3}{5}$ & -$\frac{3}{5}$ & -$1$ &
  -$\frac{\sqrt{3/2}}{10}$ &
 $\frac{1}{2\sqrt{10}}$ & $\sqrt{\frac{3}{5}}$ \\
%\hline
       & $({\bf 189}, {\bf 1})$ 3  & ${\bf 1}$ 1 & $0$ & $4\sqrt{2/5}$ &
 -$\frac{1}{2\sqrt{15}}$ & -$3$ & $0$ & $0$ & -$\sqrt{3/2}$ &
 $\sqrt{5/2}$ & $0$ \\
%\hline
  &     & ${\bf 8}$ 2 & $\frac{5}{2\sqrt{6}}$ & -$\frac{3}{2\sqrt{10}}$ &
 $\frac{\sqrt{3/5}}{2}$ &  -$1$ & $0$ & $0$ & $\frac{\sqrt{3/2}}{2}$ &
 -$\frac{\sqrt{5/2}}{2}$ & $0$  \\
%\hline
\hline
\hline
${\bf 2430}$ & $({\bf 1}, {\bf 1})$ 1& ${\bf 1}$ 1 & $0$ &
 $0$ & $\frac{13\sqrt{5/3}}{12}$ & $1$ & $\frac{101}{36}$ & $\frac{25}{12}$ &
 $0$ & $0$ & $\frac{13\sqrt{5/3}}{12}$   \\
%\hline
 & $({\bf 189}, {\bf 1})$ 2  & ${\bf 1}$ 1 & -$\frac{17\sqrt{3/2}}{2}$ &
 -$\frac{7\sqrt{5/2}}{2}$ & $\frac{\sqrt{15}}{4}$ & -$3$ & $0$ & $0$ &
 -$\sqrt{3/2}$ & $\sqrt{5/2}$ & $0$   \\
%%\hline
        &   & ${\bf 8}$ 2 & $0$ & -$2\sqrt{2/5}$ &
 -$\frac{3\sqrt{3/5}}{4}$ & -$1$ & $0$ & $0$ & $\frac{\sqrt{3/2}}{2}$ &
 -$\frac{\sqrt{5/2}}{2}$ & $0$  \\
%\hline
 & $({\bf 405}, {\bf 1})$ 3& ${\bf 1}$ 1 & $\frac{7\sqrt{3/2}}{5}$ &
 -$\frac{7}{\sqrt{10}}$ & -$\frac{3\sqrt{3/5}}{2}$ & $\frac{33}{5}$ &
 $\frac{9}{10}$ & $\frac{3}{2}$ &
 $\frac{7\sqrt{3/2}}{5}$ & -$\frac{7}{\sqrt{10}}$ & -$\frac{3\sqrt{3/5}}{2}$ \\
%\hline
 &                         & ${\bf 8}$ 2 & -$\frac{13\sqrt{3/2}}{10}$ &
 $\frac{13}{2\sqrt{10}}$ & $3\sqrt{3/5}$ & $\frac{19}{5}$ & -$\frac{9}{5}$ &
 -$3$ &
 -$\frac{13\sqrt{3/2}}{10}$ & $\frac{13}{2\sqrt{10}}$ & $3\sqrt{3/5}$ \\
%\hline
 &                        & ${\bf 27}$ 3 & -$\frac{9\sqrt{3/2}}{5}$ &
 $\frac{9}{\sqrt{10}}$ & -$\frac{9\sqrt{3/5}}{2}$ & $\frac{9}{5}$ &
 $\frac{27}{10}$ & $\frac{9}{2}$ &
 -$\frac{9\sqrt{3/2}}{5}$ & $\frac{9}{\sqrt{10}}$ & -$\frac{9\sqrt{3/5}}{2}$ \\
\hline
\end{tabular}
}
%\end{center}
%\caption{X, non-diagonal part}
\caption{
The non-diagonal part, $hij$ in Eq. (\ref{phir})
of the SM singlets $\Phi_{s,z}^r$ in each of the irreps $r$
according to their transformation properties
under the $SU(3)_L \subset SU(6)\times SU(2)_X \subset E_6 $ subgroup.
The entries $h3_{s,z}^{\prime r}$, $h5_{s,z}^{\prime r}$,
$h9_{s,z}^{\prime r}$ for the flipped embedding are also listed here.}
%\caption{The non-diagonal part, i.e., hij, in Eq. (\ref{phir}) of the standard model singlets $\Phi_{s,z}^r$ in each of the irreps $r$ (see Eq. (\ref{reps})) of $E_6$, classified according to their transformation properties under the $SU(5)\subset SO(10) \subset E_6 $ subgroup (SO(10):second column; SU(5): third column), for both the normal and flipped embedding. The entries $h3_{s,z}^{\prime r}$, $h5_{s,z}^{\prime r}$, $h9_{s,z}^{\prime r}$ for the flipped embedding are also listed here. \label{su5n}}

\end{table}

\clearpage
\begin{table} %\scriptsize
%\begin{center}
\tabcolsep=0.01cm
\footnotesize{
\begin{tabular}{c|c|c||c|c|c|c|c|c|c|c|c|c|c|c|c|c|c|c}
\hline
$E_6$ & SU(6)$\times$ & $SU(3)_R$ z &
 $h1^{r}_{s,z}$ & $h2^{r}_{s,z}$ & $h3^{r}_{s,z}$ & $h4^{r}_{s,z}$ &
 $h5^{r}_{s,z}$ & $h6^{r}_{s,z}$ & $h7^{r}_{s,z}$ & $h8^{r}_{s,z}$ &
 $h9^{r}_{s,z}$ & $h10^{r}_{s,z}$ & $h11^{r}_{s,z}$ & $h12^{r}_{s,z}$ &
 $h13^{r}_{s,z}$ & $h14^{r}_{s,z}$ & $h15^{r}_{s,z}$ & $N^r_{s,z}$  \\
 & SU(2)$_L$ s & & & & & & & & & & & & & & & & \\
\hline
\hline
${\bf 1}$ & ${\bf 1}$ 1 & ${\bf 1}$  1 &  1 & 1 & 1 & 1 & 1&  1&
 1& 1 & 1 & 1 & 1 & 1&  1& 1& 1 &$\frac{1}{\sqrt{78}}$ \\
\hline
\hline
${\bf 650}$ & $({\bf 1}, {\bf 1})$ 1& ${\bf 1}$  1 &  1 & -$5$ &
 1 & -$\frac{1}{2}$ & 1 & -$\frac{1}{2}$ & 1 & 1 & 1 & -$\frac{1}{2}$ & 1 & 1 &
 1 & -$\frac{1}{2}$ & 1 & $\frac{1}{2\sqrt{30}}$ \\
% \hline
  & $({\bf 35}, {\bf 1})$ 2& ${\bf 1}$ 1 & 1 & 0 & -$\frac{4}{5}$ & 0 & -$\frac{7}{10}$ &
 0 & 0 & -$1$ & -$\frac{1}{2}$ & 0 & 0 & -$1$ & 0 & 0 & -$1$ & $\frac{1}{4}$ \\
%\hline
       & $({\bf 189}, {\bf 1})$ 3  & ${\bf 1}$ 1 & 1 & 0 & $\frac{72}{75}$ &
 0 & -$\frac{169}{50}$ & 0 & -$\frac{2}{3}$ & 1 & -$\frac{3}{2}$ & 0 & -$\frac{2}{3}$ & 1 &
 -$\frac{2}{3}$ & 0 & 1 & $\frac{\sqrt{5/74}}{2}$ \\
%\hline
  &     & ${\bf 8}$ 2 & 0 & 0 & 0 & 0 & 0 & 0 & $\frac{5}{3}$ & -$\frac{5}{2}$ & 0 &
 0 & $\frac{5}{24}$ & $\frac{5}{4}$ & -$\frac{15}{8}$ & 0 & $\frac{5}{4}$ & $\frac{4\sqrt{3/149}}{5}$  \\
%\hline
\hline
\hline
${\bf 2430}$ & $({\bf 1}, {\bf 1})$ 1& ${\bf 1}$ 1 &  1 & $\frac{35}{9}$ &
 1 & -$\frac{7}{6}$ & 1 & -$\frac{7}{6}$ & 1 & 1 & 1 & -$\frac{7}{6}$ & 1 & 1 & 1 &
 -$\frac{7}{6}$ & 1 & $\frac{3\sqrt{3/910}}{2}$ \\
%\hline
 & $({\bf 189}, {\bf 1})$ 2  & ${\bf 1}$ 1 & 1 & 0 & -$6$ & 0 &
 $\frac{11}{2}$ & 0 & -$\frac{2}{3}$ & 1 & -$\frac{3}{2}$ & 0 & -$\frac{2}{3}$ & 1 & -$\frac{2}{3}$ &
 0 & 1 & $\sqrt{3/370}$ \\
%%\hline
        &   & ${\bf 8}$ 2 & 0 & 0 & 0 & 0 & 0 & 0 & -$\frac{5}{12}$ & $\frac{5}{4}$ &
 0 & 0 & -$\frac{5}{24}$ & -$\frac{5}{8}$ & $\frac{5}{8}$ & 0 & -$\frac{5}{8}$ & $\frac{\sqrt{6/7}}{5}$ \\
%\hline
 & $({\bf 405}, {\bf 1})$ 3& ${\bf 1}$ 1 & 1 & 0 & $\frac{12}{5}$ & 0 &
 $\frac{31}{10}$ & 0 & -$\frac{4}{3}$ & 1 & $\frac{9}{2}$ & 0 & -$\frac{4}{3}$ & 1 & -$\frac{4}{3}$ & 0 &
 1 & $\frac{1}{4\sqrt{7}}$ \\
%\hline
 &                         & ${\bf 8}$ 2 & 0 & 0 & -$2$ & 0 & 2 & 0 &
 -$\frac{5}{12}$ & $\frac{5}{4}$ & 0 & 0 & $\frac{35}{24}$ & -$\frac{5}{8}$ & -$\frac{25}{24}$ & 0 & -$\frac{5}{8}$ &
 $\frac{\sqrt{2/7}}{5}$ \\
%\hline
 &                        & ${\bf 27}$ 3 & 0 & 0 & $\frac{3}{19}$ & 0 & $\frac{23}{456}$ &
 0 & $\frac{10}{57}$ & $\frac{25}{228}$ & $\frac{135}{152}$ & 0 & -$\frac{5}{57}$ &
 -$\frac{25}{76}$ & -$\frac{5}{57}$ & 0 & -$\frac{25}{76}$ & $\frac{19\sqrt{6}}{65}$ \\
\hline
\end{tabular}
}
%\end{center}
%
%\caption{L, diagonal part}
\caption{
The diagonal part, $hj$ $(j=1,2,...,15)$ in Eq. (\ref{phir})
of the SM singlets $\Phi_{s,z}^r$ in each of the irreps $r$
according to their transformation properties
under the $SU(3)_R \subset SU(6)\times SU(2)_L \subset E_6 $ subgroup.}
%\caption{The diagonal part in Eq. (\ref{phir}) of the standard model singlets $\Phi_{s,z}^r$ in each of the irreps $r$ (see Eq. (\ref{reps})) of $E_6$, classified according to their transformation properties under the $SU(5)\subset SO(10) \subset E_6 $ subgroup (SO(10):second column; SU(5): third column  ), in the explicit version with the conventions described in the section II of the text, for the normal embedding. They agree between the normal and the flipped embedding of $G_{321}\subset E_6$, except for the $h3_{s,z}^r$, $h5_{s,z}^r$, $h9_{s,z}^r$. The entries $h3_{s,z}^{\prime r}$, $h5_{s,z}^{\prime r}$, $h9_{s,z}^{\prime r}$ for the flipped embedding are listed in Table III. $N^r_{s,z}$ is the normalization constant which makes $Tr(\Phi^{r}_{s,z} \Phi^{r}_{s,z})=1$ for each irrep r with specific s and z. (For both the normal and the flipped embedding, the $SU(5)$ here contains $SU(3)_C\times SU(2)_L$.)\label{su5}}
\end{table}
\begin{table}
%\footnotesize
%\begin{center}
\tabcolsep=0.1cm
\footnotesize{
\begin{tabular}{c||c|c||c|c|c||c|c|c|c|c|c|c|c|c|c|c|c|c}
\hline
$E_6$ & SU(6)$\times$ & SU(3)$_R$ z & $h35^{r}_{s,z}$ & $h39^{r}_{s,z}$ &
 $h59^{r}_{s,z}$ & $h3^{\prime r}_{s,z}$ & $h5^{\prime r}_{s,z}$ &
 $h9^{\prime r}_{s,z}$ & $h35^{\prime r}_{s,z}$ &
 $h39^{\prime r}_{s,z}$ & $h59^{\prime r}_{s,z}$  \\
 & SU(2)$_L$ s & & & & & & & & & & \\
\hline
\hline
${\bf 1}$ & ${\bf 1}$ 1 & ${\bf 1}$  1 & 0& 0&0 &0& 0&0 &0& 0&0  \\
\hline
\hline
${\bf 650}$ & $({\bf 1}, {\bf 1})$ 1& ${\bf 1}$  1 & 0 & 0 & 0 &
 1 & 1 & 1 & 0 & 0 & 0 \\
%\hline
  & $({\bf 35}, {\bf 1})$ 2& ${\bf 1}$ 1 & -$\frac{\sqrt{3/2}}{5}$ &
 $\frac{1}{\sqrt{10}}$ & -$\frac{\sqrt{3/5}}{2}$ & -$\frac{4}{5}$ & -$\frac{7}{10}$ & -$\frac{1}{2}$ &
 -$\frac{\sqrt{3/2}}{5}$ & $\frac{1}{\sqrt{10}}$ & -$\frac{\sqrt{3/5}}{2}$ \\
%\hline
       & $({\bf 189}, {\bf 1})$ 3  & ${\bf 1}$ 1 & $\frac{13\sqrt{3/2}}{25}$ &
 -$\frac{37}{5\sqrt{10}}$ & $\frac{17\sqrt{3/5}}{10}$ & 0 & $\frac{19}{10}$ &
 -$\frac{39}{10}$ & $\frac{11\sqrt{3/2}}{5}$ & -$\frac{7}{5\sqrt{10}}$ & $\frac{17\sqrt{3/5}}{10}$ \\
%\hline
  &     & ${\bf 8}$ 2 & -$\frac{25}{4\sqrt{6}}$ & $\frac{\sqrt{5/2}}{4}$ &
 $\frac{\sqrt{15}}{2}$ & 1 & -$\frac{1}{16}$ & -$\frac{15}{16}$ & $\frac{1}{8\sqrt{6}}$ &
 -$\frac{11\sqrt{5/2}}{8}$ & -$\frac{9\sqrt{15}}{16}$ \\
%\hline
\hline
\hline
${\bf 2430}$ & $({\bf 1}, {\bf 1})$ 1& ${\bf 1}$ 1 & 0 & 0 & 0 &
 1 & 1 & 1 & 0 & 0 & 0 \\
%\hline
 & $({\bf 189}, {\bf 1})$ 2  & ${\bf 1}$ 1 & $\sqrt{6}$ & 0 &
 $\frac{5\sqrt{5/3}}{2}$ & 0 & -$\frac{11}{2}$ & $\frac{7}{2}$ & -$\frac{3\sqrt{3/2}}{2}$ &
 -$\frac{5\sqrt{5/2}}{2}$ & $\frac{5\sqrt{5/3}}{2}$ \\
%%\hline
        &   & ${\bf 8}$ 2 & -$\frac{25}{4\sqrt{6}}$ & $\frac{\sqrt{5/2}}{4}$ &
 $\frac{\sqrt{15}}{2}$ & 1 & -$\frac{1}{16}$ & -$\frac{15}{16}$ & $\frac{1}{8\sqrt{6}}$ &
 -$\frac{11\sqrt{5/2}}{8}$ & -$\frac{9\sqrt{15}}{16}$ \\
%\hline
 & $({\bf 405}, {\bf 1})$ 3& ${\bf 1}$ 1 & -$\frac{7\sqrt{3/2}}{5}$ &
 $\frac{7}{\sqrt{10}}$ & -$\frac{7\sqrt{3/5}}{2}$ & $\frac{12}{5}$ & $\frac{31}{10}$ & $\frac{9}{2}$ &
 -$\frac{7\sqrt{3/2}}{5}$ & $\frac{7}{\sqrt{10}}$ & -$\frac{7\sqrt{3/5}}{2}$ \\
%\hline
 &                         & ${\bf 8}$ 2 & -$\frac{7\sqrt{3/2}}{4}$ &
 -$\frac{11\sqrt{5/2}}{4}$ & -$\frac{\sqrt{15}}{2}$ & 1 & -$\frac{61}{16}$ & $\frac{45}{16}$ &
 $\frac{27\sqrt{3/2}}{8}$ & $\frac{5\sqrt{5/2}}{8}$ & -$\frac{5\sqrt{15}}{16}$ \\
%\hline
 &                        & ${\bf 27}$ 3 & $\frac{49}{76\sqrt{6}}$ &
 -$\frac{11\sqrt{5/2}}{76}$ &  $\frac{11\sqrt{15}}{152}$ & 1 & $\frac{59}{456}$ &
 -$\frac{5}{152}$ & -$\frac{\sqrt{3/2}}{76}$ & $\frac{\sqrt{5/2}}{76}$ & $\frac{37\sqrt{5/3}}{152}$  \\
\hline
\end{tabular}
}
%\end{center}
%\caption{L, non-diagonal part}
\caption{
The non-diagonal part, $hij$ in Eq. (\ref{phir})
of the SM singlets $\Phi_{s,z}^r$ in each of the irreps $r$
according to their transformation properties
under the $SU(3)_R \subset SU(6)\times SU(2)_L \subset E_6 $ subgroup.
The entries $h3_{s,z}^{\prime r}$, $h5_{s,z}^{\prime r}$,
$h9_{s,z}^{\prime r}$ for the flipped embedding are also listed here.}
%\caption{The non-diagonal part, i.e., hij, in Eq. (\ref{phir}) of the standard model singlets $\Phi_{s,z}^r$ in each of the irreps $r$ (see Eq. (\ref{reps})) of $E_6$, classified according to their transformation properties under the $SU(5)\subset SO(10) \subset E_6 $ subgroup (SO(10):second column; SU(5): third column), for both the normal and flipped embedding. The entries $h3_{s,z}^{\prime r}$, $h5_{s,z}^{\prime r}$, $h9_{s,z}^{\prime r}$ for the flipped embedding are also listed here. \label{su5n}}

\end{table}
\clearpage

\clearpage
\begin{table} %\scriptsize
%\begin{center}
\tabcolsep=0.01cm
\footnotesize{
\begin{tabular}{c|c|c||c|c|c|c|c|c|c|c|c|c|c|c|c|c|c|c}
\hline
$E_6$ & SU(6)$\times$ & $SU(3)_L$ z &
 $h1^{r}_{s,z}$ & $h2^{r}_{s,z}$ & $h3^{r}_{s,z}$ & $h4^{r}_{s,z}$ &
 $h5^{r}_{s,z}$ & $h6^{r}_{s,z}$ & $h7^{r}_{s,z}$ & $h8^{r}_{s,z}$ &
 $h9^{r}_{s,z}$ & $h10^{r}_{s,z}$ & $h11^{r}_{s,z}$ & $h12^{r}_{s,z}$ &
 $h13^{r}_{s,z}$ & $h14^{r}_{s,z}$ & $h15^{r}_{s,z}$ & $N^r_{s,z}$  \\
 & SU(2)$_R$ s & & & & & & & & & & & & & & & & \\
\hline
\hline
${\bf 1}$ & ${\bf 1}$ 1 & ${\bf 1}$  1 &  1 & 1 & 1 & 1 & 1&  1& 1& 1 & 1 & 1 & 1 & 1&  1& 1& 1 &$1/\sqrt{78}$ \\
\hline
\hline
${\bf 650}$ & $({\bf 1}, {\bf 1})$ 1& ${\bf 1}$ 1 & 1 & 1 & 1 & 1 &
 -$\frac{11}{4}$ & -$\frac{1}{2}$ & -$\frac{1}{2}$ & -$\frac{1}{2}$ & -$\frac{5}{4}$ & -$\frac{1}{2}$ & -$\frac{1}{2}$ &
 -$\frac{1}{2}$ & 1 & 1 & -$5$ & $\frac{1}{2\sqrt{30}}$  \\
% \hline
  & $({\bf 35}, {\bf 1})$ 2& ${\bf 1}$ 1 & 1 & -$1$ & -$\frac{1}{5}$ & 0 &
 -$\frac{3}{10}$ & 0 & 0 & 0 & -$\frac{1}{2}$ & 0 & 0 & 0 & 0 & -$1$ & 0 & $\frac{1}{4}$ \\
%\hline
                     &     & ${\bf 8}$ 2 & 0 & 1 & $\frac{3}{5}$ & $\frac{1}{2}$ &
 -$\frac{3}{5}$ & 0 & 0 & 0 & -$1$ & 0 & 0 & 0 & -$1$ & -$\frac{1}{2}$ & 0 & $\frac{1}{4}$ \\
%\hline
  & $({\bf 35}, {\bf 3})$ 3& ${\bf 1}$ 1 & 0 & 0 & 0 & 0 &
 -$\frac{15}{16}$ & 0 & 0 & -$\frac{5}{4}$ & $\frac{15}{16}$ & $\frac{5}{24}$ & 0 & 0 & 0 &
 0 & 0 & $\frac{4\sqrt{3/11}}{5}$ \\
%\hline
                     &     & ${\bf 8}$ 2 & 0 & 0 & 0 & 0 &
 -$\frac{15}{16}$ & $\frac{5}{6}$ & 0 & 0 & $\frac{15}{16}$ & -$\frac{5}{6}$ & 0 & 0 & 0 & 0 &
 0 & $\frac{2}{5\sqrt{3}}$ \\
%\hline
       & $({\bf 189}, {\bf 1})$ 4 & ${\bf 1}$ 1 & 1 & 1 & -$3$ &
 -$\frac{2}{3}$ & -$1$ & 0 & 0 & 0 & 1 & 0 & 0 & 0 & -$\frac{2}{3}$ & 1 & 0 &
 $\frac{\sqrt{3/46}}{2}$ \\
%\hline
                            &     & ${\bf 8}$ 2 & 0 & 1 & -$1$ &
 -$\frac{1}{6}$ & 0 & 0 & 0 & 0 & 0 & 0 & 0 & 0 & $\frac{1}{3}$ & -$\frac{1}{2}$ & 0 &
 $\frac{1}{2\sqrt{2}}$ \\
%\hline
\hline
\hline
${\bf 2430}$ & $({\bf 1}, {\bf 1})$ 1& ${\bf 1}$ 1 & 1 & 1 & 1 & 1 &
 $\frac{101}{36}$ & -$\frac{7}{6}$ & -$\frac{7}{6}$ & -$\frac{7}{6}$ & $\frac{25}{12}$ & -$\frac{7}{6}$ & -$\frac{7}{6}$ &
 -$\frac{7}{6}$ & 1 & 1 & $\frac{35}{9}$ & $\frac{3\sqrt{3/910}}{2}$ \\
%\hline
 & $({\bf 1}, {\bf 5})$ 2  & ${\bf 1}$ 1 & 0 & 0 & 0 & 0 & $\frac{25}{24}$ &
 0 & 0 & 0 & $\frac{5}{8}$ & 0 & 0 & 0 & 0 & 0 & -$\frac{5}{6}$ & $\frac{\sqrt{6}}{5}$ \\
%\hline
 & $({\bf 35}, {\bf 3})$ 3  & ${\bf 1}$ 1 & 0 & 0 & 0 & 0 & -$\frac{15}{16}$ &
 0 & 0 & $\frac{25}{24}$ & $\frac{15}{16}$ & $\frac{5}{24}$ & 0 & -$\frac{55}{24}$ & 0 & 0 & 0 &
 $\frac{3\sqrt{2/11}}{5}$ \\
%\hline
                        &   & ${\bf 8}$ 2 & 0 & 0 & 0 & 0 & -$\frac{15}{16}$ &
 -$\frac{5}{24}$ & 0 & $\frac{5}{6}$ & $\frac{15}{16}$ & 0 & -$\frac{5}{36}$ & $\frac{5}{6}$ & 0 & 0 & 0 &
 $\frac{12\sqrt{3/95}}{5}$ \\
 & $({\bf 189}, {\bf 1})$ 4  & ${\bf 1}$1 & 1 & 1 & -$3$ & -$\frac{2}{3}$ &
 $\frac{15}{8}$ & 0 & 0 & 0 & -$\frac{15}{8}$ & 0 & 0 & 0 & -$\frac{2}{3}$ & 1 & 0 &
 $\frac{1}{\sqrt{115}}$ \\
%%\hline
%                        &   & ${\bf 8}$ 2 &
%%\hline
 & $({\bf 405}, {\bf 1})$ 5 & ${\bf 1}$ 1 & 1 & 1 & $\frac{33}{5}$ & -$\frac{4}{3}$ &
 $\frac{9}{10}$ & 0 & 0 & 0 & $\frac{3}{2}$ & 0 & 0 & 0 & -$\frac{4}{3}$ & 1 & 0 &
 $\frac{1}{4\sqrt{7}}$ \\
%\hline
 &                         & ${\bf 8}$ 2 & 0 & 1 & $\frac{19}{5}$ & -$\frac{5}{6}$ &
 -$\frac{9}{5}$ & 0 & 0 & 0 & -$3$ & 0 & 0 & 0 & $\frac{5}{3}$ & -$\frac{1}{2}$ & 0 &
 $\frac{1}{4\sqrt{5}}$ \\
%\hline
 &                        & ${\bf 27}$ 3 & 0 & 1 & $\frac{9}{5}$ & 0 & $\frac{27}{10}$ &
 0 & 0 & 0 & $\frac{9}{2}$ & 0 & 0 & 0 & 0 & -$3$ & 0 & $\frac{1}{2\sqrt{30}}$ \\
\hline
\end{tabular}
}
%\end{center}
%\caption{R, diagonal part}
\caption{
The diagonal part, $hj$ $(j=1,2,...,15)$ in Eq. (\ref{phir})
of the SM singlets $\Phi_{s,z}^r$ in each of the irreps $r$
according to their transformation properties
under the $SU(3)_L \subset SU(6)\times SU(2)_R \subset E_6 $ subgroup.}
%\caption{The diagonal part in Eq. (\ref{phir}) of the standard model singlets $\Phi_{s,z}^r$ in each of the irreps $r$ (see Eq. (\ref{reps})) of $E_6$, classified according to their transformation properties under the $SU(5)\subset SO(10) \subset E_6 $ subgroup (SO(10):second column; SU(5): third column  ), in the explicit version with the conventions described in the section II of the text, for the normal embedding. They agree between the normal and the flipped embedding of $G_{321}\subset E_6$, except for the $h3_{s,z}^r$, $h5_{s,z}^r$, $h9_{s,z}^r$. The entries $h3_{s,z}^{\prime r}$, $h5_{s,z}^{\prime r}$, $h9_{s,z}^{\prime r}$ for the flipped embedding are listed in Table III. $N^r_{s,z}$ is the normalization constant which makes $Tr(\Phi^{r}_{s,z} \Phi^{r}_{s,z})=1$ for each irrep r with specific s and z. (For both the normal and the flipped embedding, the $SU(5)$ here contains $SU(3)_C\times SU(2)_L$.)\label{su5}}

\end{table}

%\vskip-0.5cm

\begin{table}[b]%\scriptsize
\tabcolsep=0.1cm
\footnotesize{
\begin{tabular}{c||c|c||c|c|c||c|c|c|c|c|c|c|c|c|c|c|c|c}
\hline
$E_6$ & SU(6)$\times$ & SU(3)$_L$ z & $h35^{r}_{s,z}$ & $h39^{r}_{s,z}$ &
 $h59^{r}_{s,z}$ & $h3^{\prime r}_{s,z}$ & $h5^{\prime r}_{s,z}$ &
 $h9^{\prime r}_{s,z}$ & $h35^{\prime r}_{s,z}$ &
 $h39^{\prime r}_{s,z}$ & $h59^{\prime r}_{s,z}$  \\
 & SU(2)$_R$ s & & & & & & & & & & \\
\hline
\hline
${\bf 1}$ & ${\bf 1}$ 1 & ${\bf 1}$  1 & 0& 0&0 &0& 0&0 &0& 0&0  \\
\hline
\hline
${\bf 650}$ & $({\bf 1}, {\bf 1})$ 1& ${\bf 1}$ 1 & 0 & 0 &
 -$\frac{3\sqrt{15}}{4}$ & -$\frac{13}{5}$ & $\frac{17}{20}$ & -$\frac{5}{4}$ & $\frac{3\sqrt{3/2}}{5}$ &
 $\frac{9}{\sqrt{10}}$ & -$\frac{3\sqrt{3/5}}{4}$ \\
%\hline
  & $({\bf 35}, {\bf 1})$ 2& ${\bf 1}$ 1 & $\frac{\sqrt{3/2}}{5}$ &
 -$\frac{1}{\sqrt{10}}$ & $\frac{\sqrt{3/5}}{2}$ & -$\frac{1}{5}$ & -$\frac{3}{10}$ & -$\frac{1}{2}$ &
 $\frac{\sqrt{3/2}}{5}$ & -$\frac{1}{\sqrt{10}}$ & $\frac{\sqrt{3/5}}{2}$ \\
%\hline
                     &     & ${\bf 8}$ 2 & -$\frac{\sqrt{3/2}}{10}$ &
 $\frac{1}{2\sqrt{10}}$ & $\sqrt{3/5}$ & -$\frac{3}{5}$ & $\frac{3}{5}$ & -$1$ &
 $\frac{\sqrt{3/2}}{10}$ & -$\frac{\sqrt{5/2}}{2}$ & 0 \\
%\hline
  & $({\bf 35}, {\bf 3})$ 3& ${\bf 1}$ 1 & -$\frac{5}{8\sqrt{6}}$ &
 -$\frac{\sqrt{5/2}}{8}$ & $\frac{\sqrt{15}}{16}$ & 1 & -$\frac{1}{16}$ & -$\frac{15}{16}$ &
 $\frac{1}{8\sqrt{6}}$ & $\frac{\sqrt{5/2}}{8}$ & -$\frac{\sqrt{15}}{16}$ \\
%\hline
                     &     & ${\bf 8}$ 2 & -$\frac{5}{8\sqrt{6}}$ &
 -$\frac{\sqrt{5/2}}{8}$ & $\frac{\sqrt{15}}{16}$ & 1 & -$\frac{1}{16}$ & -$\frac{15}{16}$ &
 $\frac{1}{8\sqrt{6}}$ & $\frac{\sqrt{5/2}}{8}$ & -$\frac{\sqrt{15}}{16}$ \\
%\hline
       & $({\bf 189}, {\bf 1})$ 4  & ${\bf 1}$ 1 & $\sqrt{3/2}$ &
 $\frac{11}{\sqrt{10}}$ & $\frac{1}{\sqrt{15}}$ & -$\frac{3}{5}$ & -$\frac{7}{5}$ & -$1$ &
 -$\frac{17\sqrt{3/2}}{5}$ & $\frac{1}{\sqrt{10}}$ & $\frac{1}{\sqrt{15}}$ \\
%\hline
                             &     & ${\bf 8}$ 2 & $\frac{\sqrt{3/2}}{2}$ &
 -$\frac{\sqrt{5/2}}{2}$ & 0 & $\frac{1}{5}$ & -$\frac{6}{5}$ & 0 & $\frac{3\sqrt{3/2}}{10}$ &
 $\frac{1}{2\sqrt{10}}$ & $\sqrt{3/5}$ \\
%\hline
\hline
\hline
${\bf 2430}$ & $({\bf 1}, {\bf 1})$ 1& ${\bf 1}$ 1 & 0 & 0 &
 $\frac{13\sqrt{5/3}}{12}$ & $\frac{41}{15}$ & $\frac{193}{180}$ & $\frac{25}{12}$ &
 -$\frac{13}{15\sqrt{6}}$ & -$\frac{13}{3\sqrt{10}}$ & $\frac{13}{12\sqrt{15}}$ \\
%\hline
  & $({\bf 1}, {\bf 5})$ 2& ${\bf 1}$ 1 & 0 & 0 & $\frac{5\sqrt{5/3}}{8}$ &
 1 & $\frac{1}{24}$ & $\frac{5}{8}$ & -$\frac{1}{2\sqrt{6}}$ & -$\frac{\sqrt{5/2}}{2}$ & $\frac{\sqrt{5/3}}{8}$ \\
%\hline
  & $({\bf 35}, {\bf 3})$ 3& ${\bf 1}$ 1 & -$\frac{5}{8\sqrt{6}}$ &
 -$\frac{\sqrt{5/2}}{8}$ & $\frac{\sqrt{15}}{16}$ & 1 & -$\frac{1}{16}$ & -$\frac{15}{16}$ &
 $\frac{1}{8\sqrt{6}}$ & $\frac{\sqrt{5/2}}{8}$ & -$\frac{\sqrt{15}}{16}$ \\
%\hline
                     &     & ${\bf 8}$ 2 & -$\frac{5}{8\sqrt{6}}$ &
 -$\frac{\sqrt{5/2}}{8}$ & $\frac{\sqrt{15}}{16}$ & 1 & -$\frac{1}{16}$ & -$\frac{15}{16}$ &
 $\frac{1}{8\sqrt{6}}$ & $\frac{\sqrt{5/2}}{8}$ & -$\frac{\sqrt{15}}{16}$ \\
%\hline
 & $({\bf 189}, {\bf 1})$ 4  & ${\bf 1}$ 1 & -$\frac{19\sqrt{3/2}}{4}$ &
 -$\frac{5\sqrt{5/2}}{4}$ & -$\frac{\sqrt{15}}{8}$ & -$\frac{3}{5}$ & -$\frac{171}{40}$ & $\frac{15}{8}$ &
 $\frac{47\sqrt{3/2}}{20}$ & -$\frac{19}{4\sqrt{10}}$ & -$\frac{51\sqrt{3/5}}{8}$ \\
%%\hline
%                         &   & ${\bf 8}$ 2 &
%%\hline
 & $({\bf 405}, {\bf 1})$ 5& ${\bf 1}$ 1 & $\frac{7\sqrt{3/2}}{5}$ &
 -$\frac{7}{\sqrt{10}}$ & -$\frac{3\sqrt{3/5}}{2}$ & $\frac{9}{5}$ & $\frac{57}{10}$ & $\frac{3}{2}$ &
 $\frac{11\sqrt{3/2}}{5}$ & $\sqrt{5/2}$ & $\frac{\sqrt{15}}{2}$ \\
%\hline
 &                         & ${\bf 8}$ 2 & -$\frac{13\sqrt{3/2}}{10}$ &
 $\frac{13}{2\sqrt{10}}$ & $3\sqrt{3/5}$ & -$\frac{11}{5}$ & $\frac{21}{5}$ & -$3$ &
 -$\frac{3\sqrt{3/2}}{10}$ & -$\frac{17}{2\sqrt{10}}$ & -$2\sqrt{3/5}$ \\
%\hline
 &                        & ${\bf 27}$ 3 & -$\frac{9\sqrt{3/2}}{5}$ &
 $\frac{9}{\sqrt{10}}$ & -$\frac{9\sqrt{3/5}}{2}$ & $\frac{9}{5}$ & $\frac{27}{10}$ & $\frac{9}{2}$ &
 -$\frac{9\sqrt{3/2}}{5}$ & $\frac{9}{\sqrt{10}}$ & -$\frac{9\sqrt{3/5}}{2}$ \\
\hline
\end{tabular}
}
%\end{center}
%\caption{R, non-diagonal part}
\caption{
The non-diagonal part, $hij$ in Eq. (\ref{phir})
of the SM singlets $\Phi_{s,z}^r$ in each of the irreps $r$
according to their transformation properties
under the $SU(3)_L \subset SU(6)\times SU(2)_R \subset E_6 $ subgroup.
The entries $h3_{s,z}^{\prime r}$, $h5_{s,z}^{\prime r}$,
$h9_{s,z}^{\prime r}$ for the flipped embedding are also listed here.}
%\caption{The non-diagonal part, i.e., hij, in Eq. (\ref{phir}) of the standard model singlets $\Phi_{s,z}^r$ in each of the irreps $r$ (see Eq. (\ref{reps})) of $E_6$, classified according to their transformation properties under the $SU(5)\subset SO(10) \subset E_6 $ subgroup (SO(10):second column; SU(5): third column), for both the normal and flipped embedding. The entries $h3_{s,z}^{\prime r}$, $h5_{s,z}^{\prime r}$, $h9_{s,z}^{\prime r}$ for the flipped embedding are also listed here. \label{su5n}}
\end{table}

\clearpage

\clearpage

\begin{table}
%\scriptsize
%\begin{center}
\tabcolsep=0.03cm
\footnotesize{
\begin{tabular}{c|c||c|c|c|c|c|c|c|c|c|c|c|c|c|c|c|c}
\hline
$E_6$ & SU(3)$_L\times$ &
 $h1^{r}_{s,z}$ & $h2^{r}_{s,z}$ & $h3^{r}_{s,z}$ & $h4^{r}_{s,z}$ &
 $h5^{r}_{s,z}$ & $h6^{r}_{s,z}$ & $h7^{r}_{s,z}$ & $h8^{r}_{s,z}$ &
 $h9^{r}_{s,z}$ & $h10^{r}_{s,z}$ & $h11^{r}_{s,z}$ & $h12^{r}_{s,z}$ &
 $h13^{r}_{s,z}$ & $h14^{r}_{s,z}$ & $h15^{r}_{s,z}$ & $N^r_{s,z}$  \\
 & SU(3)$_R$ & & & & & & & & & & & & & & & \\
\hline
\hline
${\bf 1}$ & ${\bf 1}$ 1 & 1 & 1 & 1 & 1 & 1&  1& 1& 1 & 1 &
 1 & 1 & 1&  1& 1& 1 &$\frac{1}{\sqrt{78}}$ \\
\hline
\hline
${\bf 650}$ & $({\bf 1}, {\bf 1})_1$ & 0 & 1 & -$\frac{3}{5}$ & 0 & -$\frac{2}{5}$ & 0 &
 0 & -$1$ & 0 & 0 & 0 & -$1$ & 0 & 1 & -$1$ & $\frac{1}{\sqrt{19}}$ \\
% \hline
            & $({\bf 1}, {\bf 1})_2$ & 1 & 0 & -$\frac{4}{5}$ & 0 & -$\frac{7}{10}$ & 0 &
 0 & -$1$ & -$\frac{1}{2}$ & 0 & 0 & -$1$ & 0 & 0 & -$1$ & $\frac{\sqrt{3/7}}{4}$  \\
% \hline
            & $({\bf 1}, {\bf 8})$ & 0 & 0 & 1 & 0 & $\frac{7}{8}$ & 0 & 0 &
 0 & $\frac{5}{8}$ & 0 & 0 & -$\frac{5}{4}$ & 0 & 0 & 0 & $\frac{2}{5}$ \\
% \hline
            & $({\bf 8}, {\bf 1})$ & 0 & 1 & -$\frac{1}{5}$ & 0 & -$\frac{3}{10}$ & 0 &
 0 & 0 & -$\frac{1}{2}$ & 0 & 0 & 0 & 0 & -$\frac{1}{2}$ & 0 & $\frac{1}{\sqrt{5}}$ \\
% \hline
            & $({\bf 8}, {\bf 8})$ & 0 & 0 & -$\frac{1}{2}$ & -$\frac{25}{54}$ &
 -$\frac{3}{4}$ & 0 &  0 & 0 & $\frac{5}{4}$ & 0 & $\frac{25}{27}$ & 0 & 0 & 0 & 0 &
 $\frac{18\sqrt{2/281}}{5}$ \\
\hline
\hline
${\bf 2430}$ & $({\bf 1}, {\bf 1})$ & 1 & 1 & 1 & -$\frac{4}{9}$ & 1 & -$\frac{4}{9}$ &
 -$\frac{4}{9}$ & 1 & 1 & -$\frac{4}{9}$ & -$\frac{4}{9}$ & 1 & -$\frac{4}{9}$ & 1 & 1 & $\frac{\sqrt{3/26}}{2}$ \\
%\hline
             & $({\bf 1}, {\bf 8})$ & 0 & 0 & 1 & 0 & $\frac{7}{8}$ & 0 & 0 &
 -$\frac{5}{2}$ & $\frac{5}{8}$ & 0 & 0 & $\frac{5}{4}$ & 0 & 0 & 0 & $\frac{2}{5\sqrt{3}}$ \\
%\hline
             & $({\bf 8}, {\bf 1})$ & 0 & 1 & -$2$ & 0 & $\frac{31}{4}$ & 0 & 0 &
 0 & $\frac{13}{4}$ & 0 & 0 & 0 & 0 & -$3$ & 0 & $\frac{\sqrt{2}}{19}$ \\
%\hline
             & $({\bf 8}, {\bf 8})$ & 0 & 0 & -$\frac{1}{2}$ & $\frac{7}{16}$ & -$\frac{3}{4}$ &
 -$\frac{1}{8}$ & -$\frac{1}{8}$ & 0 & $\frac{5}{4}$ & -$\frac{1}{8}$ & -$\frac{1}{8}$ & 0 & -$1\frac{1}{8}$ & 0 & 0 &
 $\frac{8}{\sqrt{389}}$ \\
%\hline
             & $({\bf 1}, {\bf 27})$ & 0 & 0 & 1 & 0 & $\frac{7}{8}$ & 0 & 0 &
 $\frac{5}{4}$ & $\frac{5}{8}$ & 0 & 0 & $\frac{5}{4}$ & 0 & 0 & -$\frac{15}{4}$ & $\frac{\sqrt{2/3}}{5}$ \\
%\hline
             & $({\bf 27}, {\bf 1})$ & 0 & 1 & $\frac{69}{5}$ & 0 &
 -$\frac{93}{10}$ & 0 & 0 & 0 & $\frac{9}{2}$ & 0 & 0 & 0 & 0 & -$3$ & 0 &
 $\frac{1}{2\sqrt{105}}$ \\
\hline
\end{tabular}
}
%\end{center}
%\caption{3, diagonal part}
\caption{
The diagonal part, $hj$ $(j=1,2,...,15)$ in Eq. (\ref{phir})
of the SM singlets $\Phi_{s,z}^r$ in each of the irreps $r$
according to their transformation properties
under the $SU(3)_L \times SU(3)_R \subset E_6 $ subgroup.}
%\caption{The diagonal part in Eq. (\ref{phir}) of the standard model singlets $\Phi_{s,z}^r$ in each of the irreps $r$ (see Eq. (\ref{reps})) of $E_6$, classified according to their transformation properties under the $SU(5)\subset SO(10) \subset E_6 $ subgroup (SO(10):second column; SU(5): third column  ), in the explicit version with the conventions described in the section II of the text, for the normal embedding. They agree between the normal and the flipped embedding of $G_{321}\subset E_6$, except for the $h3_{s,z}^r$, $h5_{s,z}^r$, $h9_{s,z}^r$. The entries $h3_{s,z}^{\prime r}$, $h5_{s,z}^{\prime r}$, $h9_{s,z}^{\prime r}$ for the flipped embedding are listed in Table III. $N^r_{s,z}$ is the normalization constant which makes $Tr(\Phi^{r}_{s,z} \Phi^{r}_{s,z})=1$ for each irrep r with specific s and z. (For both the normal and the flipped embedding, the $SU(5)$ here contains $SU(3)_C\times SU(2)_L$.)\label{su5}}
\end{table}

\begin{table}
%\small
%\begin{center}
\tabcolsep=0.08cm
\footnotesize{
\begin{tabular}{c||c||c|c|c||c|c|c|c|c|c|c|c}
\hline
$E_6$ & SU(3)$_L\times$ & $h35^{r}_{s,z}$ & $h39^{r}_{s,z}$ &
 $h59^{r}_{s,z}$ & $h3^{\prime r}_{s,z}$ & $h5^{\prime r}_{s,z}$ &
 $h9^{\prime r}_{s,z}$ & $h35^{\prime r}_{s,z}$ &
 $h39^{\prime r}_{s,z}$ & $h59^{\prime r}_{s,z}$  \\
 & SU(3)$_R$ & & & & & & & & & \\
\hline
\hline
${\bf 1}$ & ${\bf 1}$ 1 &  0& 0&0 &0& 0&0 &0& 0&0  \\
\hline
\hline
${\bf 650}$ & $({\bf 1}, {\bf 1})_1$ & -$\sqrt{6}/5$ &
 $\sqrt{5/2}$ & 0 & -$3/5$ & -$1/40$ & -$3/8$ & -$23\sqrt{3/2}/20$ &
 -$1/(4\sqrt{10})$ & -$9\sqrt{3/5}/8$ \\
% \hline
            & $({\bf 1}, {\bf 1})_2$ & -$\sqrt{3/2}/5$ & -$7/\sqrt{10}$ &
 -$19/(2\sqrt{15})$ & -$4/5$ & -$17/10$ & $1/2$ & $9\sqrt{3/2}/5$ &
 $7/\sqrt{10}$ & -$1/(2\sqrt{15})$ \\
% \hline
            & $({\bf 1}, {\bf 8})$ & $\sqrt{3/2}/4$ & -$\sqrt{5/2}/4$ &
 $\sqrt{15}/8$ & 1 & $7/8$ & $5/8$ & $\sqrt{3/2}/4$ &
 -$\sqrt{5/2}/4$ & $\sqrt{15}/8$ \\
% \hline
            & $({\bf 8}, {\bf 1})$ & $\sqrt{3/2}/5$ & -$1/\sqrt{10}$ &
 $\sqrt{3/5}/2$ & -$1/5$ & -$3/10$ & -$1/2$ & $\sqrt{3/2}/5$ &
 -$1/\sqrt{10}$ & $\sqrt{3/5}/2$ \\
% \hline
            & $({\bf 8}, {\bf 8})$ & $\sqrt{3/2}/2$ & 0 & 0 & 1 &
 -$3/8$ & -$5/8$ & -$3\sqrt{3/2}/8$ & $3\sqrt{5/2}/8$ & $\sqrt{15}/8$ \\
\hline
\hline
${\bf 2430}$ & $({\bf 1}, {\bf 1})$ & 0 & 0 & 0 & 1 & 1 & 1 & 0 & 0 & 0 \\
%\hline
             & $({\bf 1}, {\bf 8})$ & $\sqrt{3/2}/4$ &
 -$\sqrt{5/2}/4$ & $\sqrt{15}/8$ & 1 & $7/8$ & $5/8$ & $\sqrt{3/2}/4$ &
 -$\sqrt{5/2}/4$ & $\sqrt{15}/8$ \\
%\hline
             & $({\bf 8}, {\bf 1})$ & $\sqrt{3/2}/2$ &
 $23/(2\sqrt{10})$ & -$23\sqrt{3/5}/4$ & -$1/5$ & -$17/40$ & $77/8$ &
 -$33\sqrt{3/2}/10$ & $17/(2\sqrt{10})$ & -$9\sqrt{3/5}/8$ \\
%\hline
             & $({\bf 8}, {\bf 8})$ & $\sqrt{3/2}/2$ & 0 & 0 & 1 &
 -$3/8$ & -$5/8$ & -$3\sqrt{3/2}/8$ & $3\sqrt{5/2}/8$ & $\sqrt{15}/8$ \\
%\hline
             & $({\bf 1}, {\bf 27})$ & $\sqrt{3/2}/4$ & -$\sqrt{5/2}/4$ &
 $\sqrt{15}/8$ & 1 & $7/8$ & $5/8$ & $\sqrt{3/2}/4$ &
 -$\sqrt{5/2}/4$ & $\sqrt{15}/8$ \\
%\hline
             & $({\bf 27}, {\bf 1})$ & -$19\sqrt{3/2}/5$ & $9/\sqrt{10}$ &
 -$9\sqrt{3/5}/2$ & $9/5$ & $279/20$ & -$27/4$ & $7\sqrt{3/2}/10$ &
 $33/(2\sqrt{10})$ & -$33\sqrt{3/5}/4$ \\
\hline
\end{tabular}
}
%\end{center}
%\caption{3, non-diagonal part}
\caption{
The non-diagonal part, $hij$ in Eq. (\ref{phir})
of the SM singlets $\Phi_{s,z}^r$ in each of the irreps $r$
according to their transformation properties
under the $SU(3)_L \times SU(3)_R \subset E_6 $ subgroup.
The entries $h3_{s,z}^{\prime r}$, $h5_{s,z}^{\prime r}$,
$h9_{s,z}^{\prime r}$ for the flipped embedding are also listed here.}
%\caption{The non-diagonal part, i.e., hij, in Eq. (\ref{phir}) of the standard model singlets $\Phi_{s,z}^r$ in each of the irreps $r$ (see Eq. (\ref{reps})) of $E_6$, classified according to their transformation properties under the $SU(5)\subset SO(10) \subset E_6 $ subgroup (SO(10):second column; SU(5): third column), for both the normal and flipped embedding. The entries $h3_{s,z}^{\prime r}$, $h5_{s,z}^{\prime r}$, $h9_{s,z}^{\prime r}$ for the flipped embedding are also listed here. \label{su5n}}
\end{table}

\section{Appendix B: structure constants of $E_6$}

In Cartan-Weyl basis, generators of a group G can be written as
$\{H_j, E_\alpha\}$, $j=1,...,l$, l=rank[G], $\alpha=\alpha^b$ is a root,
$b=1,...,(N-l)/2$ for positive root, N=order(G)=the number of generators of G.
The Lie algebra is,
\be \label{ap1} [H_j, E_\alpha] = \alpha^b_j E_\alpha,
\ee
where $\alpha^b_j$ is j-th component of $\alpha^b$ % (1.1)
and
\be \label{ap2} [E_{\alpha^i},E_{\alpha^j}]=N(\alpha^n) E_{\alpha^n},
\ee with $\alpha^n=\alpha^i+\alpha^j$ fixed.   % (1.2)
The $N(\alpha^n)$ in (\ref{ap2}) is called real SC.
We call the SC directly obtained from roots
in (\ref{ap1}) as "direct SC" for simplicity.
Then all SC of a group G are composed of real SC and direct SC.

The structure constants of $E_6$ have been
given in a Chevalley base~\cite{sc}.
We transform them into the usual form mostly used by physicists
and in the study of GUT (see, Eq. (\ref{afsc})).
Starting from the table "Structure constants for E6",
page 1526 in Vavilov's paper~\cite{sc},
we obtain the real SC of $E_6$, $f[i,j,k]$, $i,j,k=1,...,78$,
which are totally antisymmetric.
The part of f[i,j,k], which are not zero and ordered according to $i<j<k$,
is given as follows,

%\documentclass[letterpaper]{revtex4}
%\begin{document}

%\begin{eqnarray}
\allowdisplaybreaks
\footnotesize{
\begin{align} \label{frea}
&&f[1,  4,  7]=1/2;
f[1,  5,  6]=-1/2;
f[1,  13,  16]=1/2;
f[1,  14,  15]=-1/2;
f[1,  19,  22]=1/2;
f[1,  20,  21]=-1/2; \nonumber\\
&&
f[1,  26,  29]=1/2;
f[1,  27,  28]=-1/2; 
f[1,  32,  35]=1/2;
f[1,  33,  34]=-1/2;
f[1,  40,  43]=-1/2; \nonumber\\
&&
f[1,  41,  42]=1/2;
f[1,  47,  50]=1/2;
f[1,  48,  49]=-1/2;
f[1,  53,  56]=-1/2;
f[1,  54,  55]=1/2; \nonumber\\
&&
f[1,  61,  64]=1/2;
f[1,  62,  63]=-1/2;
f[1,  69,  72]=1/2;
f[1,  70,  71]=-1/2;
f[2,  4,  6]=1/2; \nonumber\\
&&
f[2,  5,  7]=1/2;
f[2,  13,  15]=1/2;
f[2,  14,  16]=1/2;
f[2,  19,  21]=1/2;
f[2,  20,  22]=1/2;
f[2,  26,  28]=1/2;\nonumber\\
&&
f[2,  27,  29]=1/2;
f[2,  32,  34]=1/2;
f[2,  33,  35]=1/2;
f[2,  40,  42]=1/2;
f[2,  41,  43]=1/2;
f[2,  47,  49]=1/2;\nonumber\\
&&
f[2,  48,  50]=1/2;
f[2,  53,  55]=-1/2;
f[2,  54,  56]=-1/2;
f[2,  61,  63]=1/2;
f[2,  62,  64]=1/2;\nonumber\\
&&
f[2,  69,  71]=1/2;
f[2,  70,  72]=1/2;
f[4,  13,  18]=-1/2;
f[4,  14,  17]=1/2;
f[4,  19,  24]=-1/2;\nonumber\\
&&
f[4,  20,  23]=1/2;
f[4,  26,  31]=-1/2;
f[4,  27,  30]=1/2;
f[4,  32,  37]=1/2;
f[4,  33,  36]=-1/2;\nonumber\\
&&
f[4,  38,  41]=-1/2;
f[4,  39,  40]=1/2;
f[4,  47,  52]=-1/2;
f[4,  48,  51]=1/2;
f[4,  53,  58]=-1/2;\nonumber\\
&&
f[4,  54,  57]=1/2;
f[4,  59,  64]=1/2;
f[4,  60,  63]=-1/2;
f[4,  67,  72]=-1/2;
f[4,  68,  71]=1/2;\nonumber\\
&&
f[5,  13,  17]=-1/2;
f[5,  14,  18]=-1/2;
f[5,  19,  23]=-1/2;
f[5,  20,  24]=-1/2;
f[5,  26,  30]=-1/2;\nonumber\\
&&
f[5,  27,  31]=-1/2;
f[5,  32,  36]=1/2;
f[5,  33,  37]=1/2;
f[5,  38,  40]=-1/2;
f[5,  39,  41]=-1/2;\nonumber\\
&&
f[5,  47,  51]=-1/2;
f[5,  48,  52]=-1/2;
f[5,  53,  57]=-1/2;
f[5,  54,  58]=-1/2;
f[5,  59,  63]=1/2;\nonumber\\
&&
f[5,  60,  64]=1/2;
f[5,  67,  71]=-1/2;
f[5,  68,  72]=-1/2;
f[6,  15,  18]=-1/2;
f[6,  16,  17]=1/2;\nonumber\\
&&
f[6,  21,  24]=-1/2;
f[6,  22,  23]=1/2;
f[6,  28,  31]=-1/2;
f[6,  29,  30]=1/2;
f[6,  34,  37]=1/2;\nonumber\\
&&
f[6,  35,  36]=-1/2;
f[6,  38,  43]=-1/2;
f[6,  39,  42]=1/2;
f[6,  49,  52]=-1/2;
f[6,  50,  51]=1/2;\nonumber\\
&&
f[6,  55,  58]=1/2;
f[6,  56,  57]=-1/2;
f[6,  59,  62]=-1/2;
f[6,  60,  61]=1/2;
f[6,  67,  70]=1/2;\nonumber\\
&&
f[6,  68,  69]=-1/2;
f[7,  15,  17]=-1/2;
f[7,  16,  18]=-1/2;
f[7,  21,  23]=-1/2;
f[7,  22,  24]=-1/2;\nonumber\\
&&
f[7,  28,  30]=-1/2;
f[7,  29,  31]=-1/2;
f[7,  34,  36]=1/2;
f[7,  35,  37]=1/2;
f[7,  38,  42]=-1/2;\nonumber\\
&&
f[7,  39,  43]=-1/2;
f[7,  49,  51]=-1/2;
f[7,  50,  52]=-1/2;
f[7,  55,  57]=1/2;
f[7,  56,  58]=1/2;\nonumber\\
&&
f[7,  59,  61]=-1/2;
f[7,  60,  62]=-1/2;
f[7,  67,  69]=1/2;
f[7,  68,  70]=1/2;
f[9,  13,  20]=-1/2;\nonumber\\
&&
f[9,  14,  19]=1/2;
f[9,  15,  22]=-1/2;
f[9,  16,  21]=1/2;
f[9,  17,  24]=-1/2;
f[9,  18,  23]=1/2;\nonumber\\
&&
f[9,  26,  33]=-1/2;
f[9,  27,  32]=1/2;
f[9,  28,  35]=-1/2;
f[9,  29,  34]=1/2;
f[9,  30,  37]=1/2;\nonumber\\
&&
f[9,  31,  36]=-1/2;
f[9,  47,  54]=1/2;
f[9,  48,  53]=-1/2;
f[9,  49,  56]=-1/2;
f[9,  50,  55]=1/2;\nonumber\\
&&
f[9,  51,  58]=1/2;
f[9,  52,  57]=-1/2;
f[9,  73,  76]=-1/2;
f[9,  74,  75]=1/2;
f[10,  13,  19]=1/2;\nonumber\\
&&
f[10,  14,  20]=1/2;
f[10,  15,  21]=1/2;
f[10,  16,  22]=1/2;
f[10,  17,  23]=1/2;
f[10,  18,  24]=1/2;\nonumber\\
&&
f[10,  26,  32]=-1/2;
f[10,  27,  33]=-1/2;
f[10,  28,  34]=-1/2;
f[10,  29,  35]=-1/2;
f[10,  30,  36]=1/2;\nonumber\\
&&
f[10,  31,  37]=1/2;
f[10,  47,  53]=1/2;
f[10,  48,  54]=1/2;
f[10,  49,  55]=-1/2;
f[10,  50,  56]=-1/2;\nonumber\\
&&
f[10,  51,  57]=1/2;
f[10,  52,  58]=1/2;
f[10,  73,  75]=-1/2;
f[10,  74,  76]=-1/2;
f[13,  28,  39]=-1/2;\nonumber\\
&&
f[13,  29,  38]=1/2;
f[13,  30,  43]=-1/2;
f[13,  31,  42]=1/2;
f[13,  32,  45]=-1/2;
f[13,  33,  44]=1/2;\nonumber\\
&&
f[13,  49,  60]=1/2;
f[13,  50,  59]=-1/2;
f[13,  51,  62]=1/2;
f[13,  52,  61]=-1/2;
f[13,  53,  66]=-1/2;\nonumber\\
&&
f[13,  54,  65]=-1/2;
f[13,  71,  76]=1/2;
f[13,  72,  75]=-1/2;
f[14,  28,  38]=1/2;
f[14,  29,  39]=1/2;\nonumber\\
&&
f[14,  30,  42]=1/2;
f[14,  31,  43]=1/2;
f[14,  32,  44]=-1/2;
f[14,  33,  45]=-1/2;
f[14,  49,  59]=-1/2;\nonumber\\
&&
f[14,  50,  60]=-1/2;
f[14,  51,  61]=-1/2;
f[14,  52,  62]=-1/2;
f[14,  53,  65]=1/2;
f[14,  54,  66]=-1/2;\nonumber\\
&&
f[14,  71,  75]=-1/2;
f[14,  72,  76]=-1/2;
f[15,  26,  39]=1/2;
f[15,  27,  38]=-1/2;
f[15,  30,  41]=1/2;\nonumber\\
&&
f[15,  31,  40]=-1/2;
f[15,  34,  45]=-1/2;
f[15,  35,  44]=1/2;
f[15,  47,  60]=-1/2;
f[15,  48,  59]=1/2;\nonumber\\
&&
f[15,  51,  64]=1/2;
f[15,  52,  63]=-1/2;
f[15,  55,  66]=1/2;
f[15,  56,  65]=1/2;
f[15,  69,  76]=-1/2;\nonumber\\
&&
f[15,  70,  75]=1/2;
f[16,  26,  38]=-1/2;
f[16,  27,  39]=-1/2;
f[16,  30,  40]=-1/2;
f[16,  31,  41]=-1/2;\nonumber\\
&&
f[16,  34,  44]=-1/2;
f[16,  35,  45]=-1/2;
f[16,  47,  59]=1/2;
f[16,  48,  60]=1/2;
f[16,  51,  63]=-1/2;\nonumber\\
&&
f[16,  52,  64]=-1/2;
f[16,  55,  65]=-1/2;
f[16,  56,  66]=1/2;
f[16,  69,  75]=1/2;
f[16,  70,  76]=1/2;\nonumber\\
&&
f[17,  26,  43]=1/2;
f[17,  27,  42]=-1/2;
f[17,  28,  41]=-1/2;
f[17,  29,  40]=1/2;
f[17,  36,  45]=1/2;\nonumber\\
&&
f[17,  37,  44]=-1/2;
f[17,  47,  62]=-1/2;
f[17,  48,  61]=1/2;
f[17,  49,  64]=-1/2;
f[17,  50,  63]=1/2;\nonumber\\
&&
f[17,  57,  66]=-1/2;
f[17,  58,  65]=-1/2;
f[17,  67,  76]=-1/2;
f[17,  68,  75]=1/2;
f[18,  26,  42]=-1/2;\nonumber\\
&&
f[18,  27,  43]=-1/2;
f[18,  28,  40]=1/2;
f[18,  29,  41]=1/2;
f[18,  36,  44]=1/2;
f[18,  37,  45]=1/2;\nonumber\\
&&
f[18,  47,  61]=1/2;
f[18,  48,  62]=1/2;
f[18,  49,  63]=1/2;
f[18,  50,  64]=1/2;
f[18,  57,  65]=1/2;\nonumber\\
&&
f[18,  58,  66]=-1/2;
f[18,  67,  75]=1/2;
f[18,  68,  76]=1/2;
f[19,  26,  45]=-1/2;
f[19,  27,  44]=1/2;\nonumber\\
&&
f[19,  34,  39]=1/2;
f[19,  35,  38]=-1/2;
f[19,  36,  43]=-1/2;
f[19,  37,  42]=1/2;
f[19,  47,  66]=1/2;\nonumber\\
&&
f[19,  48,  65]=1/2;
f[19,  55,  60]=-1/2;
f[19,  56,  59]=1/2;
f[19,  57,  62]=1/2;
f[19,  58,  61]=-1/2;\nonumber\\
&&
f[19,  71,  74]=1/2;
f[19,  72,  73]=-1/2;
f[20,  26,  44]=-1/2;
f[20,  27,  45]=-1/2;
f[20,  34,  38]=-1/2;\nonumber\\
&&
f[20,  35,  39]=-1/2;
f[20,  36,  42]=1/2;
f[20,  37,  43]=1/2;
f[20,  47,  65]=-1/2;
f[20,  48,  66]=1/2;\nonumber\\
&&
f[20,  55,  59]=1/2;
f[20,  56,  60]=1/2;
f[20,  57,  61]=-1/2;
f[20,  58,  62]=-1/2;
f[20,  71,  73]=-1/2;\nonumber\\
&&
f[20,  72,  74]=-1/2;
f[21,  28,  45]=-1/2;
f[21,  29,  44]=1/2;
f[21,  32,  39]=-1/2;
f[21,  33,  38]=1/2;\nonumber\\
&&
f[21,  36,  41]=1/2;
f[21,  37,  40]=-1/2;
f[21,  49,  66]=1/2;
f[21,  50,  65]=1/2;
f[21,  53,  60]=-1/2;\nonumber\\
&&
f[21,  54,  59]=1/2;
f[21,  57,  64]=1/2;
f[21,  58,  63]=-1/2;
f[21,  69,  74]=-1/2;
f[21,  70,  73]=1/2;\nonumber\\
&&
f[22,  28,  44]=-1/2;
f[22,  29,  45]=-1/2;
f[22,  32,  38]=1/2;
f[22,  33,  39]=1/2;
f[22,  36,  40]=-1/2;\nonumber\\
&&
f[22,  37,  41]=-1/2;
f[22,  49,  65]=-1/2;
f[22,  50,  66]=1/2;
f[22,  53,  59]=1/2;
f[22,  54,  60]=1/2;\nonumber\\
&&
f[22,  57,  63]=-1/2;
f[22,  58,  64]=-1/2;
f[22,  69,  73]=1/2;
f[22,  70,  74]=1/2;
f[23,  30,  45]=-1/2;\nonumber\\
&&
f[23,  31,  44]=1/2;
f[23,  32,  43]=-1/2;
f[23,  33,  42]=1/2;
f[23,  34,  41]=1/2;
f[23,  35,  40]=-1/2;\nonumber\\
&&
f[23,  51,  66]=1/2;
f[23,  52,  65]=1/2;
f[23,  53,  62]=-1/2;
f[23,  54,  61]=1/2;
f[23,  55,  64]=1/2;\nonumber\\
&&
f[23,  56,  63]=-1/2;
f[23,  67,  74]=-1/2;
f[23,  68,  73]=1/2;
f[24,  30,  44]=-1/2;
f[24,  31,  45]=-1/2;\nonumber\\
&&
f[24,  32,  42]=1/2;
f[24,  33,  43]=1/2;
f[24,  34,  40]=-1/2;
f[24,  35,  41]=-1/2;
f[24,  51,  65]=-1/2;\nonumber\\
&&
f[24,  52,  66]=1/2;
f[24,  53,  61]=1/2;
f[24,  54,  62]=1/2;
f[24,  55,  63]=-1/2;
f[24,  56,  64]=-1/2;\nonumber\\
&&
f[24,  67,  73]=1/2;
f[24,  68,  74]=1/2;
f[26,  47,  78]=1/2;
f[26,  48,  77]=1/2;
f[26,  55,  68]=-1/2;\nonumber\\
&&
f[26,  56,  67]=1/2;
f[26,  57,  70]=-1/2;
f[26,  58,  69]=1/2;
f[26,  63,  74]=1/2;
f[26,  64,  73]=-1/2;\nonumber\\
&&
f[27,  47,  77]=-1/2;
f[27,  48,  78]=1/2;
f[27,  55,  67]=1/2;
f[27,  56,  68]=1/2;
f[27,  57,  69]=1/2;\nonumber\\
&&
f[27,  58,  70]=1/2;
f[27,  63,  73]=-1/2;
f[27,  64,  74]=-1/2;
f[28,  49,  78]=1/2;
f[28,  50,  77]=1/2;\nonumber\\
&&
f[28,  53,  68]=-1/2;
f[28,  54,  67]=1/2;
f[28,  57,  72]=-1/2;
f[28,  58,  71]=1/2;
f[28,  61,  74]=-1/2;\nonumber\\
&&
f[28,  62,  73]=1/2;
f[29,  49,  77]=-1/2;
f[29,  50,  78]=1/2;
f[29,  53,  67]=1/2;
f[29,  54,  68]=1/2;\nonumber\\
&&
f[29,  57,  71]=1/2;
f[29,  58,  72]=1/2;
f[29,  61,  73]=1/2;
f[29,  62,  74]=1/2;
f[30,  51,  78]=1/2;\nonumber\\
&&
f[30,  52,  77]=1/2;
f[30,  53,  70]=1/2;
f[30,  54,  69]=-1/2;
f[30,  55,  72]=-1/2;
f[30,  56,  71]=1/2;\nonumber\\
&&
f[30,  59,  74]=1/2;
f[30,  60,  73]=-1/2;
f[31,  51,  77]=-1/2;
f[31,  52,  78]=1/2;
f[31,  53,  69]=-1/2;\nonumber\\
&&
f[31,  54,  70]=-1/2;
f[31,  55,  71]=1/2;
f[31,  56,  72]=1/2;
f[31,  59,  73]=-1/2;
f[31,  60,  74]=-1/2;\nonumber\\
&&
f[32,  49,  68]=1/2;
f[32,  50,  67]=-1/2;
f[32,  51,  70]=-1/2;
f[32,  52,  69]=1/2;
f[32,  53,  78]=-1/2;\nonumber\\
&&
f[32,  54,  77]=-1/2;
f[32,  63,  76]=1/2;
f[32,  64,  75]=-1/2;
f[33,  49,  67]=-1/2;
f[33,  50,  68]=-1/2;\nonumber\\
&&
f[33,  51,  69]=1/2;
f[33,  52,  70]=1/2;
f[33,  53,  77]=1/2;
f[33,  54,  78]=-1/2;
f[33,  63,  75]=-1/2;\nonumber\\
&&
f[33,  64,  76]=-1/2;
f[34,  47,  68]=-1/2;
f[34,  48,  67]=1/2;
f[34,  51,  72]=-1/2;
f[34,  52,  71]=1/2;\nonumber\\
&&
f[34,  55,  78]=1/2;
f[34,  56,  77]=1/2;
f[34,  61,  76]=-1/2;
f[34,  62,  75]=1/2;
f[35,  47,  67]=1/2;\nonumber\\
&&
f[35,  48,  68]=1/2;
f[35,  51,  71]=1/2;
f[35,  52,  72]=1/2;
f[35,  55,  77]=-1/2;
f[35,  56,  78]=1/2;\nonumber\\
&&
f[35,  61,  75]=1/2;
f[35,  62,  76]=1/2;
f[36,  47,  70]=-1/2;
f[36,  48,  69]=1/2;
f[36,  49,  72]=-1/2;\nonumber\\
&&
f[36,  50,  71]=1/2;
f[36,  57,  78]=1/2;
f[36,  58,  77]=1/2;
f[36,  59,  76]=-1/2;
f[36,  60,  75]=1/2;\nonumber\\
&&
f[37,  47,  69]=1/2;
f[37,  48,  70]=1/2;
f[37,  49,  71]=1/2;
f[37,  50,  72]=1/2;
f[37,  57,  77]=-1/2;\nonumber\\
&&
f[37,  58,  78]=1/2;
f[37,  59,  75]=1/2;
f[37,  60,  76]=1/2;
f[38,  51,  74]=-1/2;
f[38,  52,  73]=1/2;\nonumber\\
&&
f[38,  57,  76]=1/2;
f[38,  58,  75]=-1/2;
f[38,  59,  78]=-1/2;
f[38,  60,  77]=-1/2;
f[38,  65,  68]=-1/2;\nonumber\\
&&
f[38,  66,  67]=-1/2;
f[39,  51,  73]=1/2;
f[39,  52,  74]=1/2;
f[39,  57,  75]=-1/2;
f[39,  58,  76]=-1/2;\nonumber\\
&&
f[39,  59,  77]=1/2;
f[39,  60,  78]=-1/2;
f[39,  65,  67]=1/2;
f[39,  66,  68]=-1/2;
f[40,  47,  74]=1/2;\nonumber\\
&&
f[40,  48,  73]=-1/2;
f[40,  53,  76]=-1/2;
f[40,  54,  75]=1/2;
f[40,  63,  78]=1/2;
f[40,  64,  77]=1/2;\nonumber\\
&&
f[40,  65,  72]=-1/2;
f[40,  66,  71]=-1/2;
f[41,  47,  73]=-1/2;
f[41,  48,  74]=-1/2;
f[41,  53,  75]=1/2;\nonumber\\
&&
f[41,  54,  76]=1/2;
f[41,  63,  77]=-1/2;
f[41,  64,  78]=1/2;
f[41,  65,  71]=1/2;
f[41,  66,  72]=-1/2;\nonumber\\
&&
f[42,  49,  74]=1/2;
f[42,  50,  73]=-1/2;
f[42,  55,  76]=1/2;
f[42,  56,  75]=-1/2;
f[42,  61,  78]=-1/2;\nonumber\\
&&
f[42,  62,  77]=-1/2;
f[42,  65,  70]=1/2;
f[42,  66,  69]=1/2;
f[43,  49,  73]=-1/2;
f[43,  50,  74]=-1/2;\nonumber\\
&&
f[43,  55,  75]=-1/2;
f[43,  56,  76]=-1/2;
f[43,  61,  77]=1/2;
f[43,  62,  78]=-1/2;
f[43,  65,  69]=-1/2;\nonumber\\
&&
f[43,  66,  70]=1/2;
f[44,  59,  68]=1/2;
f[44,  60,  67]=-1/2;
f[44,  61,  70]=-1/2;
f[44,  62,  69]=1/2;\nonumber\\
&&
f[44,  63,  72]=-1/2;
f[44,  64,  71]=1/2;
f[44,  65,  78]=1/2;
f[44,  66,  77]=-1/2;
f[45,  59,  67]=-1/2;\nonumber\\
&&
f[45,  60,  68]=-1/2;
f[45,  61,  69]=1/2;
f[45,  62,  70]=1/2;
f[45,  63,  71]=1/2;
f[45,  64,  72]=1/2;\nonumber\\
&&
f[45,  65,  77]=-1/2;
f[45,  66,  78]=-1/2 .
%\end{eqnarray}
\end{align}
}

%\end{document}

%\newpage
\normalsize
The other parts of $f[i,j,k]$ either can be obtained
from Eq. (\ref{frea}) by using their antisymmetric properties,
or are equal to zero.

The positive roots of $E_6$ in Dynkin basis have been given
(see, Table 20 in Ref.~\cite{Slansky:1981yr}).
It is straightforward to get the direct SC, $fsc[i,j,k]$, $i,j,k=1,...,78$.
The part of $fsc[i,j,k]$, which are not zero and
ordered according to $i<j<k$, is given in Eq. (\ref{fdisc}).
The other parts of $fsc[i,j,k]$ can either be obtained
from Eq. (\ref{fdisc}) by using their antisymmetric properties,
or are equal to zero.

%\documentclass[letterpaper]{revtex4}
%\begin{document}

\allowdisplaybreaks
{\footnotesize
\begin{align}
%\begin{eqnarray}
\label{fdisc}
&&fsc[1,2,3]=1;
fsc[3,4,5]=\frac{1}{2};
fsc[3,6,7]=-\frac{1}{2};
fsc[3,13,14]=\frac{1}{2};
fsc[3,15,16]=-\frac{1}{2}; \nonumber\\
&&
fsc[3,19,20]=\frac{1}{2};
fsc[3,21,22]=-\frac{1}{2};
fsc[3,26,27]=\frac{1}{2};
fsc[3,28,29]=-\frac{1}{2};
fsc[3,32,33]=\frac{1}{2}; \nonumber\\
&&
fsc[3,34,35]=-\frac{1}{2};
fsc[3,40,41]=-\frac{1}{2};
fsc[3,42,43]=\frac{1}{2};
fsc[3,47,48]=\frac{1}{2};
fsc[3,49,50]=-\frac{1}{2};\nonumber\\
&&
fsc[3,53,54]=\frac{1}{2};
fsc[3,55,56]=-\frac{1}{2};
fsc[3,61,62]=\frac{1}{2};
fsc[3,63,64]=-\frac{1}{2};
fsc[3,69,70]=\frac{1}{2};\nonumber\\
&&
fsc[3,71,72]=-\frac{1}{2};
fsc[4,5,8]=\frac{\sqrt{3}}{2};
fsc[6,7,8]=\frac{\sqrt{3}}{2};
fsc[8,13,14]=\frac{1}{2\sqrt{3}};
fsc[8,15,16]=\frac{1}{2\sqrt{3}};\nonumber\\
&&
fsc[8,17,18]=-\frac{1}{\sqrt{3}};
fsc[8,19,20]=\frac{1}{2\sqrt{3}};
fsc[8,21,22]=\frac{1}{2\sqrt{3}};
fsc[8,23,24]=-\frac{1}{\sqrt{3}}; \nonumber\\
&&
fsc[8,26,27]=\frac{1}{2\sqrt{3}};
fsc[8,28,29]=\frac{1}{2\sqrt{3}};
fsc[8,30,31]=-\frac{1}{\sqrt{3}};
fsc[8,32,33]=\frac{1}{2\sqrt{3}};\nonumber\\
&&
fsc[8,34,35]=\frac{1}{2\sqrt{3}};
fsc[8,36,37]=-\frac{1}{\sqrt{3}};
fsc[8,38,39]=\frac{1}{\sqrt{3}};
fsc[8,40,41]=-\frac{1}{2\sqrt{3}};\nonumber\\
&&
fsc[8,42,43]=-\frac{1}{2\sqrt{3}};
fsc[8,47,48]=\frac{1}{2\sqrt{3}};
fsc[8,49,50]=\frac{1}{2\sqrt{3}};
fsc[8,51,52]=-\frac{1}{\sqrt{3}};\nonumber\\
&&
fsc[8,53,54]=\frac{1}{2\sqrt{3}};
fsc[8,55,56]=\frac{1}{2\sqrt{3}};
fsc[8,57,58]=-\frac{1}{\sqrt{3}};
fsc[8,59,60]=\frac{1}{\sqrt{3}};\nonumber\\
&&
fsc[8,61,62]=-\frac{1}{2\sqrt{3}};
fsc[8,63,64]=-\frac{1}{2\sqrt{3}};
fsc[8,67,68]=\frac{1}{\sqrt{3}};
fsc[8,69,70]=-\frac{1}{2\sqrt{3}};\nonumber\\
&&
fsc[8,71,72]=-\frac{1}{2\sqrt{3}};
fsc[9,10,11]=1;
fsc[11,13,14]=-\frac{1}{2};
fsc[11,15,16]=-\frac{1}{2};
fsc[11,17,18]=-\frac{1}{2};\nonumber\\
&&
fsc[11,19,20]=\frac{1}{2};
fsc[11,21,22]=\frac{1}{2};
fsc[11,23,24]=\frac{1}{2};
fsc[11,26,27]=\frac{1}{2};
fsc[11,28,29]=\frac{1}{2};\nonumber\\
&&
fsc[11,30,31]=\frac{1}{2};
fsc[11,32,33]=-\frac{1}{2};
fsc[11,34,35]=-\frac{1}{2};
fsc[11,36,37]=-\frac{1}{2};
fsc[11,47,48]=\frac{1}{2};\nonumber\\
&&
fsc[11,49,50]=\frac{1}{2};
fsc[11,51,52]=\frac{1}{2};
fsc[11,53,54]=-\frac{1}{2};
fsc[11,55,56]=-\frac{1}{2};
fsc[11,57,58]=-\frac{1}{2};\nonumber\\
&&
fsc[11,73,74]=\frac{1}{2};
fsc[11,75,76]=-\frac{1}{2};
fsc[12,13,14]=-\frac{\sqrt{\frac{5}{3}}}{2};
fsc[12,15,16]=-\frac{\sqrt{\frac{5}{3}}}{2};\nonumber\\
&&
fsc[12,17,18]=-\frac{\sqrt{\frac{5}{3}}}{2};
fsc[12,19,20]=-\frac{\sqrt{\frac{5}{3}}}{2};
fsc[12,21,22]=-\frac{\sqrt{\frac{5}{3}}}{2};
fsc[12,23,24]=-\frac{\sqrt{\frac{5}{3}}}{2};\nonumber\\
&&
fsc[12,26,27]=\frac{1}{2\sqrt{15}};
fsc[12,28,29]=\frac{1}{2\sqrt{15}};
fsc[12,30,31]=\frac{1}{2\sqrt{15}};
fsc[12,32,33]=\frac{1}{2\sqrt{15}};\nonumber\\
&&
fsc[12,34,35]=\frac{1}{2\sqrt{15}};
fsc[12,36,37]=\frac{1}{2\sqrt{15}};
fsc[12,38,39]=-\frac{2}{\sqrt{15}};
fsc[12,40,41]=-\frac{2}{\sqrt{15}};\nonumber\\
&&
fsc[12,42,43]=-\frac{2}{\sqrt{15}};
fsc[12,44,45]=\sqrt{\frac{3}{5}};
fsc[12,47,48]=\frac{1}{2\sqrt{15}};
fsc[12,49,50]=\frac{1}{2\sqrt{15}};\nonumber\\
&&
fsc[12,51,52]=\frac{1}{2\sqrt{15}};
fsc[12,53,54]=\frac{1}{2\sqrt{15}};
fsc[12,55,56]=\frac{1}{2\sqrt{15}};
fsc[12,57,58]=\frac{1}{2\sqrt{15}};\nonumber\\
&&
fsc[12,59,60]=-\frac{2}{\sqrt{15}};
fsc[12,61,62]=-\frac{2}{\sqrt{15}};
fsc[12,63,64]=-\frac{2}{\sqrt{15}};
fsc[12,65,66]=-\sqrt{\frac{3}{5}};\nonumber\\
&&
fsc[12,67,68]=\frac{1}{\sqrt{15}};
fsc[12,69,70]=\frac{1}{\sqrt{15}};
fsc[12,71,72]=\frac{1}{\sqrt{15}};
fsc[12,73,74]=-\frac{\sqrt{\frac{3}{5}}}{2};\nonumber\\
&&
fsc[12,75,76]=-\frac{\sqrt{\frac{3}{5}}}{2};
fsc[25,26,27]=\sqrt{\frac{2}{5}};
fsc[25,28,29]=\sqrt{\frac{2}{5}};
fsc[25,30,31]=\sqrt{\frac{2}{5}};\nonumber\\
&&
fsc[25,32,33]=\sqrt{\frac{2}{5}};
fsc[25,34,35]=\sqrt{\frac{2}{5}};
fsc[25,36,37]=\sqrt{\frac{2}{5}};
fsc[25,38,39]=\sqrt{\frac{2}{5}};\nonumber\\
&&
fsc[25,40,41]=\sqrt{\frac{2}{5}};
fsc[25,42,43]=\sqrt{\frac{2}{5}};
fsc[25,44,45]=\sqrt{\frac{2}{5}};
fsc[25,47,48]=-\frac{1}{2\sqrt{10}};\nonumber\\
&&
fsc[25,49,50]=-\frac{1}{2\sqrt{10}};
fsc[25,51,52]=-\frac{1}{2\sqrt{10}};
fsc[25,53,54]=-\frac{1}{2\sqrt{10}};
fsc[25,55,56]=-\frac{1}{2\sqrt{10}};\nonumber\\
&&
fsc[25,57,58]=-\frac{1}{2\sqrt{10}};
fsc[25,59,60]=-\frac{1}{2\sqrt{10}};
fsc[25,61,62]=-\frac{1}{2\sqrt{10}};
fsc[25,63,64]=-\frac{1}{2\sqrt{10}};\nonumber\\
&&
fsc[25,65,66]=\frac{1}{2\sqrt{10}};
fsc[25,67,68]=\frac{3}{2\sqrt{10}};
fsc[25,69,70]=\frac{3}{2\sqrt{10}};
fsc[25,71,72]=\frac{3}{2\sqrt{10}};\nonumber\\
&&
fsc[25,73,74]=\frac{3}{2\sqrt{10}};
fsc[25,75,76]=\frac{3}{2\sqrt{10}};
fsc[25,77,78]=\frac{\sqrt{\frac{5}{2}}}{2};
fsc[46,47,48]=-\frac{\sqrt{\frac{3}{2}}}{2};\nonumber\\
&&
fsc[46,49,50]=-\frac{\sqrt{\frac{3}{2}}}{2};
fsc[46,51,52]=-\frac{\sqrt{\frac{3}{2}}}{2};
fsc[46,53,54]=-\frac{\sqrt{\frac{3}{2}}}{2};
fsc[46,55,56]=-\frac{\sqrt{\frac{3}{2}}}{2};\nonumber\\
&&
fsc[46,57,58]=-\frac{\sqrt{\frac{3}{2}}}{2};
fsc[46,59,60]=-\frac{\sqrt{\frac{3}{2}}}{2};
fsc[46,61,62]=-\frac{\sqrt{\frac{3}{2}}}{2};
fsc[46,63,64]=-\frac{\sqrt{\frac{3}{2}}}{2};\nonumber\\
&&
fsc[46,65,66]=\frac{\sqrt{\frac{3}{2}}}{2};
fsc[46,67,68]=-\frac{\sqrt{\frac{3}{2}}}{2};
fsc[46,69,70]=-\frac{\sqrt{\frac{3}{2}}}{2};
fsc[46,71,72]=-\frac{\sqrt{\frac{3}{2}}}{2};\nonumber\\
&&
fsc[46,73,74]=-\frac{\sqrt{\frac{3}{2}}}{2};
fsc[46,75,76]=-\frac{\sqrt{\frac{3}{2}}}{2};
fsc[46,77,78]=\frac{\sqrt{\frac{3}{2}}}{2} .
\end{align}
%\end{eqnarray}
}

%\end{document}

As said above, all SC of $E_6$, $ff[i,j,k]$\footnote{In Eq. (\ref{afsc}),
all SC of a group G is denoted by $f_{ijk}$.
We now denote all SC of $E_6$ by $ff[i,j,k]$.},
are the sum of real SC and direct SC.
That is, $ff[i,j,k]=f[i,j,k]+fsc[i,j,k]$.
It has been checked that $ff[i,j,k]$ satisfy the Jacobi identities.

\section*{Acknowledgments}

This research was supported in part by the Natural Science Foundation of
China under grant numbers 11375248, 11647601, 11135009 and 11005033.

\end{document}